               \font\frm=cmr10
\def\0{\over } \def\6{\partial }
\def\({\left(} \def\){\right)}  \def\<{\langle } \def\>{\rangle }
\def\lk{\,\left[ \,} \def\rk{\,\right] \,}
\def\lb{\left\{} \def\rb{\right\}} 
\def\lsim{\,\lower 4pt\hbox{$\sim\!\!\!\!\!$}\raise 2pt\hbox{$<$}\,}
\def\gsim{\,\lower 4pt\hbox{$\sim\!\!\!\!\!$}\raise 2pt\hbox{$>$}\,}
\def\mn{ _{\mu \nu} }        \def\omn{ ^{\mu \nu} }
\def\be{ \begin{equation} }  \def\bea{ \begin{eqnarray} }
\def\ee{ \end{equation} }    \def\eea{ \end{eqnarray} }
\def\nonu{ \nonumber }       \def\eff{{\rm eff}}
\def\sgn{\,\mbox{sgn}\,}     \def\wsc{\overline{\o}}
\def\dis{ \displaystyle }    \def\Tr{ \, {\rm Tr} \, }
\def\parag#1{ \vspace{1.5cm} \hspace{.08cm} \parbox{15cm}{{\bf #1}
       \vspace{1.1cm} } \hfill \vphantom{a} \nopagebreak \indent }
\def\wu#1{\sqrt{{#1} \,}^{ \hbox to0.2pt{\hss$ 
        \vrule height 2pt width 0.6pt depth 0pt $} \;\! } }
\def\w{\omega_k}  \def\wp{\omega_k+i\zeta}  \def\wm{\omega_k-i\zeta}
\def\P{ {\mit\Pi} }           \def\X{ {\mit\Xi} }
\let\a=\alpha  \let\g=\gamma  \let\d=\delta \let\e=\varepsilon
\let\z=\zeta   \let\th=\theta \let\s=\sigma \let\o=\omega
\let\O=\Omega  \let\D=\Delta  \let\G=\Gamma
\let\thq=\theequation        
   \def\pfeil{_\rightharpoonup}   \def\leer{\phantom{a}}
   \def\opf{\buildrel \pfeil \over \leer}
   \def\jvv{j \lower0.4pt\hbox to 2pt{\hss $\opf$}}
   \def\jv{j \lower0.2pt\hbox to 1.4pt{\hss $\opf$}}
   \def\ivv{i \lower0.4pt\hbox to 2pt{\hss $\opf$}}
   \def\iv{i \lower0.2pt\hbox to 1.4pt{\hss $\opf$}}
   \def\hq{h \raise0.2pt\hbox to 0.4pt{\hss $^-$}}
   \def\vk#1{\hbox{$\buildrel           \pfeil \over #1$}}
   \def\vkk#1{\hbox{$\buildrel   \;     \pfeil \over #1$}}
   \def\vkkk#1{\hbox{$\buildrel  \, \;  \pfeil \over #1$}}
   \def\grpf{\displaystyle  _\rightharpoonup}
   \def\vg#1{\hbox{$\buildrel       \grpf \over #1$}}
   \def\vgg#1{\hbox{$\buildrel  \;  \grpf \over #1$}}
\def\fzz{f} \def\bzz{b} \def\dzz{d} \def\gzz{g} \def\hzz{h}
\def\jzz{j} \def\kzz{k} \def\lzz{l} \def\mzz{m} \def\wzz{w}
\def\tzz{t} \def\izz{i} \def\bezz{\beta} \def\dezz{\delta}
\def\xizz{\xi} \def\pszz{\psi} \def\vthzz{\vartheta}
\def\uph{ \! \mathop{\vphantom{a}} } \def\dph{ \vphantom{a} }
\def\vc#1{\def\tast{\noexpand#1} \def\test{#1}
    \ifcat\tast\bzz
      \ifx\test\fzz \vkkk f \uph \else
       \ifx\test\bzz \vkk b \uph \else
        \ifx\test\dzz \vkkk d \uph \else
         \ifx\test\gzz \vkk g \dph \else
          \ifx\test\hzz \vkk h \uph \else
           \ifx\test\izz \ivv \dph \else
            \ifx\test\jzz \jvv \dph \else
             \ifx\test\kzz \vkk k \uph \else
              \ifx\test\lzz \vkk l \uph \else
               \ifx\test\tzz \vkk t \uph \else
                \ifx\test\mzz \vg m \dph \else
                 \ifx\test\wzz \vg w \dph \else
                  \ifnum \lq#1<91 \vgg #1 \uph \else \vk #1 \dph
                  \fi
                 \fi
                \fi
               \fi
              \fi
             \fi
            \fi
           \fi
          \fi
         \fi
        \fi
       \fi
      \fi
    \else
     \ifx\test\bezz \vkk \beta \uph \else
      \ifx\test\pszz \vkk \psi \dph \else
       \ifx\test\dezz \vkk \delta \uph \else
        \ifx\test\xizz \vkk \xi \uph \else
         \ifx\test\vthzz \vkk \vartheta \uph \else \vk #1 \dph
         \fi
        \fi
       \fi
      \fi
     \fi
    \fi }  
   \documentstyle[12pt,twoside%
]{article} 

\begin{document}
%
%
\begin{titlepage}
   \noindent{\large\tt DESY 94-034 \hfill ISSN 0418-9833  \\
   ITP-UH-01/94 \hfill hep-ph/9403301 \\
   March 1994  }     \vfill  \vskip 2.5cm
\begin{center}    \vfill \bigskip \bigskip
 {\Large \bf   Resummations in Hot Scalar Electrodynamics }
     \vskip 1.6cm \vfill \vfill
 {\large U. Kraemmer , \ A. K. Rebhan$^1$ \ and \ H. Schulz$^2$ }\\
\end{center}
\medskip \smallskip
\qquad \qquad {\sl $^1$ DESY, Theory Group, Notkestr. 
               85, D-22603 Hamburg, Germany} \\ \medskip
\qquad \qquad {\sl $^2$} \parbox[t]{12cm}{ \sl 
    Institut f\"ur Theoretische Physik, Universit\"at Hannover, \\ 
    Appelstr. 2, D-30167 Hannover, Germany }  \vskip 2.5cm \vfill
%
%
  \hspace{5cm} {\large  ABSTRACT}  \begin{quotation}
The gauge-boson sector of perturbative scalar electrodynamics is
investigated in detail as a testing ground for resummation methods
in hot gauge theories. It also serves as a simple non-trivial
reference system for the non-Abelian gluon plasma. The complete
next-to-leading order contributions to the polarization tensor are
obtained within the resummation scheme of Braaten and Pisarski. The
simpler scheme proposed recently by Arnold and Espinosa is shown to
apply to static quantities only, whereas Braaten-Pisarski
resummation turns out to need modification for collective phenomena
close to the light-cone. Finally, a recently proposed resummation of
quasi-particle damping contributions is assessed critically.
  \end{quotation}  \end{titlepage}

%
%
\let\dq=\thq \renewcommand{\theequation}{1.\dq}
\setcounter{equation}{0}

\parag {1. \ Introduction }

Perturbative thermal gauge field theories at ultra-relativistic
temperatures \cite{Kapu} are hoped to be applicable for the
description of phenomena associated with the envisaged formation
of a quark-gluon plasma in high-energy ion collisions and for the
physics of the early universe. In recent years, a lot of work has
been put into developing a systematic perturbation scheme, which
has turned out to require a resummation of the usual series of
loop diagrams.

In non-Abelian gauge theories it was found by Linde \cite{Linde}
that there exists a barrier raised by infrared divergences
associated with unscreened static magnetic fields, but even before
this barrier is hit, conventional perturbation theory breaks down
for certain mass scales much smaller than the temperature. In hot
quantum chromodynamics (QCD), which is the fundamental theory
relevant for the hypothetical quark-gluon plasma, it is for example
the spectrum of quasi-particles whose perturbative treatment
requires a resummed perturbation theory. After some failed attempts
to calculate dissipative properties of gluonic plasma excitations
\cite{samm}, such a resummation scheme was developed finally by
Braaten and Pisarski \cite{BP}. Also in spontaneously broken
theories, there has been a flurry of activity centered around the
issue of determining corrections to the effective potentials of
scalar fields which describe the cosmological phase transition from
the high-temperature symmetric to the low-temperature broken phase.
These developments, which have been pioneered in the early seventies
\cite{DJ}, are summarized e.g.~in Ref.~\cite{AE}.

The need for a resummation of the ordinary loop expansion is not
peculiar to non-Abelian theories, but can be studied also in the
Abelian case. One of the simplest models is scalar electrodynamics.
As Higgs model, it has already been used extensively to investigate
the nature of the phase transition at finite temperature, e.g.~in
Ref.~\cite{BHW}. There the scalar sector is of central interest. In
this paper we shall concern ourselves with the sector of the gauge
fields, and we shall use hot scalar electrodynamics to study {\it en
miniature} the issues that have been raised in the case of hot QCD.
Scalar electrodynamics appears to be a potentially interesting toy
model for purely gluonic QCD, since it involves self-interacting
bosons and a gauge-field sector. In hot QCD, the resummed
perturbation theory of Ref.~\cite{BP} has been employed to derive a
consistent, gauge-independent damping constant for the lowest-energy
plasmon mode \cite{Damping} as well as for energetic quasi-particles
\cite{BMAR}, and also some next-to-leading order results:
corrections to the plasma frequency \cite{HS} and to Debye screening
\cite{AKR}. In the much simpler case of hot scalar electrodynamics,
we shall be able to give a fairly complete picture for the photonic
quasi-particle spectrum.

In doing so, we shall largely follow the resummation program of
Braaten and Pisarski \cite{BP}. After presenting the leading-order
results in Sect.~2, this scheme is reviewed for the simple case of
scalar electrodynamics in Sect.~3. In Sect.~4 we first specialize
to the static limit and obtain next-to-leading order results
for Debye screening and magnetic permeability. Here the simpler
resummation scheme of Ref.~\cite{AE}, which has originally been put
forward for the resummation of the static effective potential,
agrees with full resummation. In Sect.~5, the limit of
long-wavelength oscillations is considered and the next-to-leading
order result for the plasma frequency is found. In this case, a
naive application of the resummation scheme of Arnold and Espinosa
is shown to fail, because nonstatic modes can no longer be
neglected. We also show to what extent classical considerations can
explain the value of the plasma frequency obtained before. In
Sect.~6 we exhibit the complete next-to-leading order results for
the polarization tensor, deriving the corrected spectrum of
propagating photonic quasi-particles. Here we find a qualitative
change in the case of the longitudinal plasmons, which turn out to
exist only for a finite range of frequencies or momenta. The
canonical resummation according to Braaten and Pisarski actually
breaks down in this example due to singularities at the light-cone,
but can be amended so that this upper bound for the frequencies of
longitudinal plasmons can be calculated accurately. A similar result
is shown to hold in the case of QCD. Finally, in Sect.~7 we consider
the additional resummation of scalar damping contributions,
critically assessing a recent proposal in Ref.~\cite{MC}. Sect.~8
contains our conclusions.

%
%
\let\dq=\thq \renewcommand{\theequation}{2.\dq}
\setcounter{equation}{0}

\parag {2. \ Leading-order results }  

Scalar electrodynamics with a scalar potential leading to spontaneous
symmetry breaking has been extensively used as a toy model to study
symmetry restoration at high temperature and the corresponding phase
transition which occurs when the temperature is lowered \cite{BHW}.
In this paper we shall focus our attention to the gauge field sector
of this model rather than the effective potential of the scalars.
For this we shall for simplicity consider charged scalar particles
without self-interactions. At sufficiently high temperature, the bare
mass of the scalar particles can be neglected, so our starting point
is the Lagrangian \cite{IZ}
\be  \label{2lagr}
  {\cal L} \, = \, \( D_\mu \varphi \) ^\ast D^\mu \varphi
     - {1 \0 4} \, F\mn F\omn
     - {1 \0 2 \a} \,\( \6^\mu \! A_\mu \)^2
     \;\; ,
\ee
with the covariant derivative $D_\mu=\6_\mu + i e A_\mu$ and
the Abelian field tensor $F\mn = \6_\mu A_\nu - \6_\nu A_\mu$. The
signature of the Minkowski metric $g\mn$ is $+---$. In (\ref{2lagr})
we have included a gauge breaking term corresponding to general
covariant gauges; the ghost term is omitted since it decouples from
the rest.

We shall mainly use the Matsubara formalism, where Green's functions
are first defined for discrete imaginary frequencies proportional to
$2\pi iT$. After the evaluation of all frequency sums, they are
finally extended to real external frequencies by an appropriate
analytic continuation. In the first chapters we shall have to
consider exclusively the continuation of two-point functions, for
which we choose retarded boundary conditions prescribing
$Q_0 = 2\pi i n T \to \o +i \e$ \cite{Kapu,LW}.

The theory (\ref{2lagr}) has two propagators and two vertices.
Writing a four-momentum as $Q=(i\o_n, \vc q)$,
$\o_n=2\pi n T$, the bare scalar propagator is $S^0=-1/Q^2$,
and the bare photon propagator reads $G^0\mn=g\mn /Q^2+(\a -1)
D\mn /Q^2$. Here $D$ is the fourth matrix of the following basis for
symmetric Lorentz tensors,
\be  \label{2AD}  \hspace{-.5cm}
  A = g-B-D \;\; , \;\; B= { V \circ V \0 V^2 } \;\; , \;\;
  C = { Q \circ V + V \circ Q \0 \wu 2 Q^2 q } \;\; , \;\;
  D = { Q \circ Q \0 Q^2 } \;\; ,
\ee
where $V \equiv Q^2 U - (UQ) \, Q $ is a projector to
longitudinal fields built from the rest-frame velocity of
the heat bath $U$, which in the following we choose as
$U=(1\, ,\vc 0\,)$. $A$ and $B$ are transverse with respect to
$Q_\mu$; $A$ is transverse also with respect to the three-momentum
$\vc q$. The translation into the notation of the textbook of
Kapusta \cite{Kapu} is given by $A=-P_T$ and $B=-P_L$. The vertex
which couples a scalar line ($Q$ ingoing, $P$ outgoing) with one
photon is $-e(Q+P)^\mu$, and that connecting to two photon lines
is $2e^2g\omn$ with the Lorentz indices carried by the two photons.

\def\do{{$\!$.}}                                                 
\begin{figure}[t] \unitlength.8cm \begin{picture}(18.7,4.3)
\multiput(1,3)(.137,0){16}{\do}     \put(2,3.04){\circle*{.3}}
  \put(3.8,2.9){$=$} 
\multiput(5,3)(.137,0){12}{\do}     \put(7.3,2.9){$+$}
\multiput(8.5,3)(.137,0){7}{\do}    \put(10,3.1){\circle{1.44}}
\put(11.7,3.04){\circle*{.3}}  
\multiput(10.8,3)(.137,0){15}{\do}  \put(13.5,2.9){$+$}
\multiput(14.7,3)(.137,0){29}{\do}  \put(16,3.72){\circle{1.44}}
\put(17.5,3.04){\circle*{.3}}
\put(1,.5){\line(1,0){2}}  \put(2,.5){\circle*{.3}}   
  \put(3.8,.4){$=$} 
\put(5,.5){\line(1,0){1.5}}    \put(7.3,.4){$+$}
\put(8.5,.5){\line(1,0){4.14}} \put(11.7,.5){\circle*{.3}}
  \put(13.5,.4){$+$}
\put(14.7,.5){\line(1,0){3.74}} \put(17.5,.5){\circle*{.3}}
 \put(16.700,1.200)\do  \put(16.000,1.900)\do  \put(15.300,1.200)\do
 \put(16.687,1.337)\do  \put(15.863,1.887)\do  \put(15.313,1.063)\do
 \put(16.647,1.468)\do  \put(15.732,1.847)\do  \put(15.353,.932)\do
 \put(16.582,1.589)\do  \put(15.611,1.782)\do  \put(15.418,.811)\do
 \put(16.495,1.695)\do  \put(15.505,1.695)\do  \put(15.505,.705)\do
 \put(16.389,1.782)\do  \put(15.418,1.589)\do  \put(15.611,.618)\do
 \put(16.268,1.847)\do  \put(15.353,1.468)\do  \put(15.732,.553)\do
 \put(16.137,1.887)\do  \put(15.313,1.337)\do  \put(15.863,.513)\do
 \put(16.000,.500)\do   \put(16.137,.513)\do   \put(16.268,.553)\do
 \put(16.389,.618)\do   \put(16.495,.705)\do   \put(16.582,.811)\do
 \put(16.647,.932)\do   \put(16.687,1.063)\do
 \put(10.700,.500)\do  \put(10.000,1.200)\do  \put(9.300,.500)\do
 \put(10.687,.637)\do  \put(9.863,1.187)\do
 \put(10.647,.768)\do  \put(9.732,1.147)\do
 \put(10.582,.889)\do  \put(9.611,1.082)\do
 \put(10.495,.995)\do  \put(9.505,.995)\do
 \put(10.389,1.082)\do \put(9.418,.889)\do
 \put(10.268,1.147)\do \put(9.353,.768)\do
 \put(10.137,1.187)\do \put(9.313,.637)\do
\end{picture}    
\caption[f1]{\label{f1}{\frm Propagators dressed by hard
   thermal loops. Dotted lines represent photons, and
   solid lines stand for scalar particles.}}                     
\end{figure}
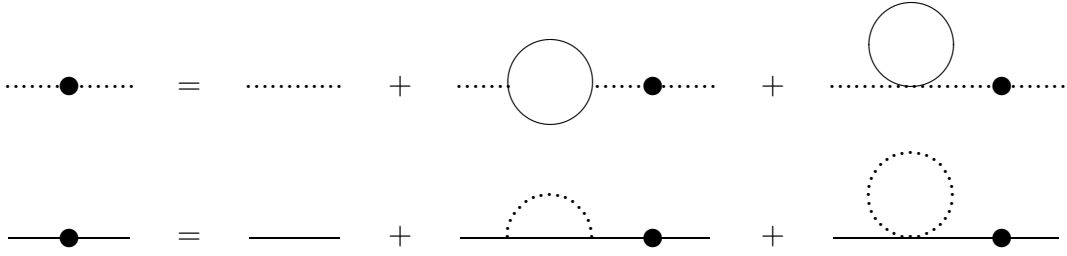                                                     

At one-loop level, the propagators are given by Dyson's equation
\be   \label{2GS}
  G = G^0+G^0 \P G \qquad \mbox{and} \qquad S = S^0+S^0 \X S \;\; ,
\ee
where $\P$ and $\X$ are the self energies of photon and scalar,
respectively. The corresponding Feynman diagrams are depicted in
Fig.~1.

The bare 1-loop expression for the photon self-energy reads
\be   \label{2Pi}
\P\omn (Q) = {3\0 2} m^2 g\omn +e^2 \sum {4 K^\mu K^\nu -Q^\mu Q^\nu
  \0 K^2 (K-Q)^2 } \quad \mbox{with} \quad m \equiv {eT \0 3} \;\; ,
\ee
where the blank sum symbol stands for $(2\pi )^{-3} \!\int\! d^3k
\,\, T \sum_n$. We shall always use $K$ as loop-momentum variable.
In contrast to the self-energy of gluons in QCD, (\ref{2Pi}) is
gauge parameter independent, and also somewhat simpler. $\P\omn$ is
transverse, $Q_\mu \P\omn = 0$ which in Abelian theories holds for
any linear gauge choice (in non-Abelian theories the
finite-temperature gluon self-energy is, in general, non-transverse
even in covariant gauges). It therefore decomposes into
\be
  \P\omn = A\omn \P_t+B\omn \P_\ell
\ee
with $\P_t = \Tr A\P /2$ and $\P_\ell = \Tr B\P$.

In the high-temperature limit, the leading order contributions are
given by
\bea \label{plhtl}
  \P_\ell & = & -{Q^2 \0 q^2} \P_{00} = 3m^2 \( 1 - {Q_0^2 \0 q^2}
  \) \(1 - {Q_0 \0 2q} \ln{Q_0+q\0Q_0-q}\) \;\; , \\
 \label{pthtl}
    \P_t & = & {1\02}\(3m^2-\P_\ell\) \;\; .
\eea

For the comparison with subleading results to be discussed later on,
in Fig.~2 the real part of $\P_\ell$ is displayed as a function of
real $Q_0=\o$ and $q$ (according to (\ref{pthtl}) $\P_t$ is
simply given by the inverted picture, appropriately relabelled).
Notice in particular the non-analytic behavior at the origin, where
the value of $\P_\ell$ (and also of $\P_t$) depends on whether it
is approached from space-like, time-like, or light-like directions.

The scalar self-energy on the other hand is given by
\be  \label{2Xi}
  \X = - \mu^2 - e^2 Q^2 \sum \( {3-\a \0 K^2 (K-Q)^2}
     + { 2(\a -1) KQ \0 K^4 (K-Q)^2 } \)
     \quad \mbox{with} \quad \mu \equiv {eT \0 2} \;\; .
\ee
(Including $\varphi^4$-self-interactions by ${\cal L}\to {\cal L}
-{\lambda\0 4}(\varphi^*\varphi)^2$ would only change
$\mu^2 \to e^2T^2/4 + \lambda T^2/12$.) In contrast to the photon
self-energy, $\X$ does depend on the gauge parameter $\a$, but the
only contribution $\propto T^2$ is the first term, which is again
gauge-independent.

Including only the leading order contributions $\propto e^2T^2$,
the scalar propagator according to (\ref{2GS}) is simply
\be  \label{2S}
   S = {-1\0 Q^2-\mu^2} \;\; ,
\ee
with a constant (thermal) mass term. The photon propagator, however,
has a richer structure, and reads in the tensorial basis (\ref{2AD})
\be   \label{2G}
 G\omn (Q) = A\omn {1 \0 Q^2-\P_t(Q)}
    + B\omn {1 \0 Q^2-\P_\ell (Q)} + D\omn {\a \0 Q^2} \;\; ,
\ee
with $\P_t$ and $\P_\ell$ at leading order as given by (\ref{plhtl},
\ref{pthtl}).

\begin{figure}                                                   
\caption[f2]{\label{f2}{\frm The real part of
   $\P_\ell(\o ,q)$ in units of $e^2T^2$. Also given is
   the intersection with the surface $\o^2-q^2$ which
   gives the dispersion curve for longitudinal plasmons.}}       
\end{figure}                                                     

The ultra-relativistic photon propagator (\ref{2G}) has been first
derived within classical kinetic theory by Silin \cite{Silin}, and
within field theory by Fradkin \cite{Fradkin}. Exactly the same
expressions were later found in high-temperature QCD
\cite{KalKl,Weldon}, the only difference being a replacement
$e^2 \to g^2(N+N_f/2)$ for SU($N$) with $N_f$ flavors. There are
now two physical structure functions in the gauge-field propagator.
The one associated with the spatially transverse tensor $A$
corresponds to transverse photons (or gluons), whereas the one
associated with $B$ describes a new collective mode, the so-called
plasmon mode. The poles of these structure functions correspond to
the normal modes of the gauge field sector of the ultra-relativistic
plasma. For real frequencies and momenta, these give the spectrum
of propagating photonic quasi-particles. It starts at the minimal
(plasma) frequency $\o (q=0) = eT/3 \equiv m$. For larger
frequencies the two branches differ. At $q \to \infty$ the
spatially transverse branch approaches the form $\wu {q^2 +
m_\infty^2}$ with an asymptotic thermal mass $m_\infty = eT/\wu 6$,
whereas the longitudinal branch approaches the light-cone
exponentially. This longitudinal branch is also shown in Fig.~2.
It disappears from the spectrum eventually because with increasing
momentum the residue of the corresponding pole in (\ref{2G})
vanishes exponentially. Perturbing the plasma with frequencies
$\o < m$ does not give rise to propagating quasi-particles, but
results in dynamical screening, both in the spatially transverse
and in the longitudinal mode. In the static limit, there is
screening only in the longitudinal sector corresponding to Debye
screening of longitudinal electric fields with screening mass
$m_{el.}=eT/\wu3$, whereas static magnetic fields remain unscreened.

The main theme of this work will be to determine this spectrum of
the photonic quasi-particles and the screening properties beyond
the level of the leading temperature contributions.
 
%
%
\let\dq=\thq \renewcommand{\theequation}{3.\dq}    
\setcounter{equation}{0}          

\parag {3. \ Resummation of hard thermal loops }

In perturbative quantum field theories it sometimes happens that
higher-order loop diagrams are not suppressed by correspondingly
high powers of the coupling constants, and that therefore the 
perturbation series has to be reorganized in order to actually
be perturbative. In thermal field theories this is a ubiquitious
phenomenon. In particular the zero modes of massless Bose particles
are sensitive to the appearance of thermal masses. Repeated
self-energy insertions cause higher-order diagrams to become more
and more infra-red singular, but summing them up according to
Fig.~1 the resulting dressed propagators are well-behaved.
In the case of the thermodynamic potential, this particular
resummation is known as ``ring resummation'' and dates back
to the work of Gell-Mann and Brueckner \cite{GMB}.

In the case of the spectrum of quasi-particle excitations, the
necessity of a resummed perturbation theory has become apparent
in gauge theories by the failure of the unimproved perturbation
theory to give gauge-independent results for the damping constants
of thermal quasi-particle excitations \cite{samm}. A systematic
perturbation theory was developed most notably by Braaten and
Pisarski \cite{BP}. These authors have shown that Green's functions
involving soft external momenta (i.e.~of the same order of magnitude
as the plasma frequency) require the resummation of all so-called
hard thermal loops (HTL's) \cite{FT,BP}. The latter are the
contributions from one-loop diagrams that are dominated by hard
loop momenta $\sim T$. In renormalizable theories they go like
$T^2$ for $T\to\infty$. They can compensate for powers of the
coupling as follows. Consider in particular the dressed propagators
as given by Fig.~1. In the diagrams on the right-hand-side of
Fig.~1 the self-energy part contributes a factor $\sim e^2T^2$.
If the bare propagator attached to it carries momentum $\sim eT$,
it brings in a factor $e^{-2}T^{-2}$ and the contribution involving
the self-energy insertion is of relative order 1. Only for larger
external momentum the perturbative nature of Fig.~1 is restored,
and loop-corrections are suppressed by factors of $e$. Consequently,
contributions from loop integrals (in any higher order) that probe
these soft scales are not perturbatively stable, unless (at least)
these hard thermal loops are resummed. The hard thermal loops
themselves are stable, because they are sensitive to hard scales
only.

In non-Abelian gauge theories, the same situation occurs for vertex 
functions, because there the one-loop vertices receive also
contributions $\sim T^2$, such that HTL vertices are of the same
order of magnitude as bare ones for soft external momenta.
Therefore, they have to be included in the resummation as well.

In scalar electrodynamics, there are no HTL vertices. Superficial
power counting \cite{BP} rules out (as in QCD) all candidates for
HTL vertices, which contain a bare 4-vertex or different kinds of
lines in the loop. Moreover, scalar loops with an odd number of
external lines are easily shown to vanish, because they change sign
by reversing momentum flows. For the remaining scalar loops with an
even number $n$ of external photons with momenta $Q_j$ one arrives
at
\be  \label{3n}
 e^n \sum {K_{\mu_1}K_{\mu_2} \cdots K_{\mu_n} \0
 (K+Q_1)^2 (K+Q_1+Q_2)^2 \cdots (K-Q_n)^2 K^2 }
 \;\; + \;\; {\rm perm. \; of\; Q_1,\ldots Q_{n-1}} \;\; . 
\ee
with $Q_1 + Q_2 + \ldots + Q_n = 0 $.
The power-counting arguments of Ref.~\cite{BP} would estimate
that (\ref{3n}) goes like $T^2$ for $T\to\infty$, so that with
$Q_j \sim eT$ (\ref{3n}) would become of the order
$e^{n-2}|Q|^{4-n}$, the same order of magnitude as bare vertices
are on dimensional grounds. However, for $n=4$ it has been shown in
Ref.~\cite{FT} that the leading-order contributions in (\ref{3n})
in fact cancel, and in Ref.~\cite{BPward} it was proved by
induction that this holds for all $n>4$ as well.

        \def\eff{{\rm eff}}
In non-Abelian gauge theories with or without fermions, the
HTL vertices are essential to guarantee the gauge invariance
of the generating functional of all HTL's. An explicit effective
action $S_\eff$ has been constructed by Taylor and Wong \cite{eff} 
by starting from its bilinear part $S^{(2)}_\eff$ as determined 
by the HTL self-energy diagrams and determining higher orders in 
the fields by gauge invariance. A nice representation of 
this can be found in Ref.~\cite{shortcut}. In scalar electrodynamics 
it turns out that $S^{(2)}_\eff$ is already gauge invariant so that
$S_\eff=S^{(2)}_\eff$, and the uniqueness of this construction, which
is presented in some detail in App.~A, explains the absence of HTL 
vertices in the Abelian case.

For soft momenta, the HTL's are of the same order as the tree-level
terms, so together they represent the 'zeroth order' of the
high-temperature limit of the theory. Thus, the effective action,
which combines them into a single formula, represents the first term
of the high-temperatur asymptotics of this theory. In its manifest
gauge invariant form, it reads
\be  \label{3eff}
  S_\eff = S + \d S = \int^\beta {\cal L}_{\rm eff} \quad , \quad
    {\cal L}_{\rm eff} = {\cal L} - \mu^2 \varphi^\ast \varphi
    + {3 \0 4} m^2 \int_\O Y^\rho F_{\rho \mu} \,
    { 1 \0 (Y \6 )^2 } \, F^{\mu \lambda} Y_\lambda \quad,
\ee
where $\int^\beta \equiv \int_0^\beta \! d\tau \int\! d^3r$
and $Y^\mu \equiv (1,\vc e)$ with $\vc e^2=1$. The angular integral
$\int_\O$ over the directions of $\vc e $ is normalized to one:
$\int_\O 1=1$, $\int_\O Y=U$.

When adopting (\ref{3eff}) as the adequate starting point
for a perturbative treatment, one must however take 
care of not to change the underlying theory itself. One is therefore
led to rewrite $S$ as $S=S_\eff - \d S$, where $-\d S$ generates
counter-terms that subtract at higher loop orders what has been added in
at lower ones.

The first contributions beyond the new zeroth order are contained 
in 1-loop diagrams, but now with resummed internal lines. For the
photon self-energy this resummation amounts to insert the massive
scalar propagators (\ref{2S}) into the diagrams of Fig.1. The
general form of the photon Green's function is still given by
(\ref{2G}) by virtue of gauge invariance, i.e. transversality.

The resummed 1PI expressions $\P_\ell=\Tr B \P$ and $\P_t = \Tr A
\P /2$ derive from
\be  \label{2Psubmn}
  \P\omn = -2e^2 g\omn \sum \D + 4e^2 \sum K^\mu K^\nu \D^- \D
  \;\; , \;\; \D \equiv {1 \0 K^2-\mu^2}
  \; , \; \D^- \equiv {1 \0 (K-Q)^2 -\mu^2} \;\; ,
\ee
and are given more explicitly by
\be \label{2Psubel}
  \P_\ell = - 2 e^2 \sum \D + 4 e^2 \sum \D^- \D
  \lk  \vc k ^2 - { ( \vc k \vc q )^2 \0 q^2} + K^2 
  - {(KQ)^2 \0 Q^2} \rk \;\; ,
\ee
\be \label {2Psubt}
   \P_t \; = \; {1\0 2} \; \( \; \P_g - \P_\ell \; \)  \;\; ,
\ee 
where $\P_t$ is determined to the slightly simpler auxiliary
quantity
\be  \label{2Psubg}
    \P_g \equiv \Tr g \P  = -4 e^2 \sum \D
    + e^2 (4\mu^2-Q^2) \sum \D^- \D \;\; .
\ee
In (\ref{2Psubmn}) a term involving $Q^\mu Q^\nu$ has been dropped,
because it does not contribute at the subleading order under study.

An alternative version of $\P_\ell$ is
\be \label{2Psubell}  
  \P_\ell = e^2 (4\mu^2-Q^2) \sum \D^- \D
      + 4 e^2  \sum \D^- \D 
      \lk  k^2- {(\vc k \vc q )^2 \0 q^2} \rk \;\; .
\ee
Special elements of $\P\omn$ may be read off from
$\P\omn = A\omn \P_t + B\omn \P_\ell$. In particular,
\be  \label{2ii}
 \P_{00} = - {q^2 \0 Q^2} \P_\ell \quad , \quad
 \P_{ii} = - 2\P_t - {Q_0^2 \0 Q^2} \P_\ell = \P_{00} -\P_g \;\; . 
\ee

%
%
\let\dq=\thq \renewcommand{\theequation}{4.\dq}
\setcounter{equation}{0}

\parag {4. \  Static screening at next-to-leading order  }

In the static limit the spatially-transverse and
spatially-longitudinal structure functions in the photon propagator,
as given by (\ref{2ii}), simplify to
\be
  \P_t(0,q) = - {1\0 2} \P_{ii}(0,q) \;\; , \qquad
  \P_\ell(0,q) = \P_{00}(0,q) \;\; .
\ee

Before resummation, the leading and subleading terms in the
high-temperature expansion of the static one-loop self-energy
(\ref{2Pi}) read
\be \label{p00unr}
  \P_{00}(0,q) = {e^2T^2\03} + {e^2\024\pi^2} q^2 \ln{\s\0T}
    + \ldots = 3m^2 \(1 + O(e^2) \)
\ee
and
\be \label{piiunr}
 \P_{ii}(0,q) = -{1\08}e^2qT - {e^2\012\pi^2}q^2\ln{\s\0T}
     + \ldots = -{3\08}mqe \( 1 + O(e) \) \;\; ,
\ee
where here and in what follows we count orders in $e$ always for
soft momenta $\sim m \propto eT$. Notice further that
the $T=0$ part has been assumed to be renormalized at the
scale $\s$, which contributes a term proportional to $\ln \s^2/Q^2$.
This has been combined with a similar term proportional to
$\ln Q^2/T^2$ from the temperature-dependent parts.

The leading contribution in $\P_{00}$ is the familiar electric
(Debye) screening mass $m_{\rm el.}^2=3m^2$, while there is no
screening mass of this order of magnitude in $\P_{ii}$, i.e. no
magnetic mass. At relative order $e$ there is no contribution in
$\P_{00}$, but one in $\P_{ii}$ which when taken seriously spells
trouble. Including the latter in the static transverse propagator
would give
\be
  \D_t(0,q) = {-1\0q^2-e^2qT/16} \;\; ,
\ee
which has a space-like pole at $q=e^2T/16$. A similar behaviour is
found also in QCD, sometimes called the Landau ghost of thermal QCD.

However, as we have argued in the previous sections, the subleading
terms in (\ref{p00unr}) and (\ref{piiunr}) are not accurate
for $q\lsim eT$ and require resummation of the hard thermal loops.
Only the terms proportional to $e^2\ln(\s/T)$ are to be trusted as
they are determined by the ultraviolet part of the loop integrals.

\vspace{1cm}
\noindent 
\parbox{16cm}{4.\,A \ RESUMMATION \vspace{.7cm} }
\hfill \vphantom{a} \nopagebreak \indent
$\P_{00}(0,q)$ upon resummation of the hard thermal loop, which is
just the thermal scalar mass $\mu$, reads
\be
 \P_{00}(0,q) = e^2\sum \lb 4 K_0^2 \D \D^- -2 \D \rb \;\; ,
\ee
see (\ref{2Psubel}), with $\D$ and $\D^-$ as defined in
(\ref{2Psubmn}). Evaluating the sum over the Matsubara frequencies
by means of a contour integral leads to
\be
  \P_{00}(0,q) = {e^2\0\pi^2} \int_0^\infty \! dk \, k^2 \,
    {n(\wu{k^2+\mu^2}) \0 \wu{k^2+\mu^2}} \lk 1 +
    {k^2+\mu^2\0kq}\ln\left|{2k+q\02k-q}\right| \, \rk \;\; .
\ee

After separating off the $T^2$-contribution, we can write
\be   \hspace{-.2cm}
  \P_{00}(0,q) = {e^2T^2\03} - {e^2T\0\pi^2} \int_0^\infty \! dk \,
  \lb { \mu^2 \0 k^2+\mu^2 } + 1 - {k\0q} \ln
  \left| {2k+q\02k-q} \right| \rb + O(e^2 q^2 \ln(T)) \;\; . \;\;
\ee
This can be evaluated by integration by parts, which reveals that
the $q$-dependence of the contribution at order $T$ is completely
spurious:
\be \label{p00r}
  \P_{00}(0,q) = {e^2T^2\03} - {e^2T\mu\02\pi} + O(e^2 q^2 \ln(T))
     = 3m^2 \( 1-{3\04\pi}e + O(e^2) \) \;\; .
\ee

The static $\P_{00}$ at next-to-leading order is a negative
constant, resulting in a decrease of the classical value of the
electric screening mass. In QCD the corresponding calculation has
been performed recently by one of the present authors in
Ref.~\cite{AKR}. There the next-to-leading order correction to
$\P_{00}(0,q)$ turns out to be a nontrivial function of $q$, which
is such that it diverges logarithmically for $q^2\to-m_{\rm el.}^2$,
where $\P_{00}(0,q)$ defines the correction term to the Debye mass.
Assuming that this divergence is cut-off by a magnetic mass, which
is expected to arise non-perturbatively at order $g^2T$ ($g$ being
the QCD coupling constant), leads to $\d m_{\rm el.}^2 =
O(g\ln(1/g)m^2)$ such that $\d m_{\rm el.}>0$ for small $g$.
Evidently, in this quantity there is a qualitative difference
between an Abelian and a non-Abelian gauge field plasma beyond
leading order.

In a similar manner, using (\ref{2Psubt}) and (\ref{2Psubg}), one
can evaluate the resummed expression for $\P_{ii}(0,q)$,
\be
  \P_{ii}(0,q) = e^2 \sum \lb \(4k^2-q^2\)\D\D^- + {6\D} \rb \;\; .
\ee
Its high-temperature limit is found to be
\be \label{piir}
   \P_{ii}(0,q) = {e\mu\02\pi} \lk 2\mu-{q^2+4\mu^2\0q}
   \arctan\(q\02\mu\) \rk + O(e^2 \mu^2) \;\; ,
\ee
which is to be compared with (\ref{piiunr}). The latter is only
accurate for $q\gg eT$, and in this regime it indeed coincides with
the resummed result (\ref{piir}), since $\arctan(q/(2\mu))\to\pi/2$
for $q/\mu\to\infty$.

For $q \lsim \, eT$, the resummed result deviates considerably from
the bare one. In particular for $q\to0$, the former approaches zero
like $q^2$ rather than $q$, so that there is no longer any unphysical
pole at space-like momentum. The vanishing of $\P_{ii}(0,q\to0)$
also implies that there is no magnetic mass squared of the order of
$e^3T^2$. In fact, in the present Abelian case it can be shown
rigorously that there is no magnetic screening mass
\cite{Fradkin,KalKl}. In QCD \cite{KalKl} the situation is again
quite different. Resummation changes the unphysical pole at
space-like momenta, but does not remove it. However, since this pole
arises at momentum scale $g^2T$, this just points to the relevance
of a magnetic mass in the non-Abelian case.

For finite $q\sim eT$, $\P_{ii}(0,q)$ is always negative, which
implies that the magnetic permeability
\be
  \({1\0\mu}-1\)_{\rm static} = -{1\02q^2}\P_{ii}(0,q)
\ee
is positive, decreasing monotoneously from $e/( 12\pi)$ at $q=0$
to zero for $q\to\infty$. Hence, the hot scalar plasma is weakly
diamagnetic at distances $\gsim \, 1/(eT)$.

\vspace{1cm}
\noindent  
\parbox{16cm}{4.\,B \ STATIC RESUMMATION \vspace{.7cm} }
\hfill \vphantom{a} \nopagebreak \indent
Up to now we have strictly followed the resummation scheme outlined
by Braaten and Pisarski for hot gauge theories (a detailed
exposition in the simpler scalar case can be found in
Ref.~\cite{Parwani}). However, the above calculations can be
somewhat simplified by the following observations due to Arnold and
Espinosa \cite{AE} made in the context of a resummed perturbation
theory for the finite temperature effective potential in gauge
theories. In the Matsubara formalism, $K^2$ has Euclidean signature,
$-K^2=(2\pi nT)^2+k^2$, so that the momentum $K_\mu$ can be soft
only with $n=0$. Accordingly it is sufficient to dress only these
static modes:
\be
  \D_{n=0}(k) = {-1\0 k^2+\P(0,k)} \;\; ,
\ee
whereas the correction terms to the nonstatic propagators in
\be
  \D_{n\not=0}(k) = {-1\0(2\pi nT)^2+k^2}\lk 1+\sum_{m=1}^\infty
  {(-1)^m \P^m(2\pi inT,k)\0[(2\pi nT)^2 +  k^2]^{m}} \rk
\ee
are down by powers of $g^{2m}$ when $\P\sim g^2T^2$.

A systematic resummation scheme based on this splitting has been put
forward in Ref.~\cite{AE}, where the decisive simplifications for
gauge theories are due to the fact that the hard thermal loop
$\P(0,k)$ is a constant mass term. In its lowest order version, it
just coincides with the well-known ring resummation introduced by
Gell-Mann and Brueckner \cite{GMB}.

The limitations of this restricted resummation scheme become
apparent when one considers nonstatic Green's functions. Because of
energy conservation at vertices, external frequencies are also fixed
to Matsubara frequencies, and because of the special treatment of
the static sector, it is quite impossible to perform an analytic
continuation to nonzero soft external frequencies in the end.
However, this scheme should be sufficient for the evaluation of
corrections to static quantities like the effective potential
\cite{AE} or screening masses \cite{AKR}, and to Green's functions
with hard external frequencies, e.g. damping of energetic particles
\cite{BMAR}.

Let us exemplify this simplified resummation method by recalculating
the next-to-leading order screening mass in hot scalar
electrodynamics. Separating static from nonstatic modes in
$\P_{00}(0,k)$ gives
\bea \label{split}  \hspace{-.3cm}
  \P_{00}(0,q) & = & e^2\sum \lb 4 K_0^2 \D \D^- -2 \D \rb \nonu\\
  & = & e^2T\s^{2\e}
    \int{d^{3-2\e}k\0(2\pi)^{3-2\e}}{2\0k^2+\mu^2} \nonu\\
  & & {} + 2e^2T\s^{2\e}\sum_{n=1}^\infty
  \int{d^{3-2\e}k\0(2\pi)^{3-2\e}}\lb
  -4[(k-q)^2+\mu^2]\D_n\D_n^-+6\D_n\rb \;\; . \;\;
\eea
Here we have employed dimensional regularization in order to render
the splitted expressions well-defined. From (\ref{split}) it is
apparent that indeed only the $n=0$ contributions are capable of
producing a relative order $e$ through $e^2T/\mu$ --- the other,
nonstatic contributions can be expanded out in powers of $\mu^2$ and
will not give something nonanalytic in $\mu^2$, 
neither i.e., in $e^2$.

The next-to-leading order term is thus contained in
\bea
  \d \P_{00}(0,q) & = & 2e^2T\s^{2\e} \int{d^{3-2\e}k\0 (2\pi)^{3
   - 2\e}}{1\0 k^2+\mu^2} + O(e^2 q^2 \ln T)  \nonu\\
  & = & 2e^2T\mu {\G (- {1\0 2} + \e)\0 (4\pi)^{3/2}}
        \( {4\pi\s^2\0 \mu^2} \)^\e + O(e^2 q^2 \ln T)   \nonu\\
  & = & -{e^2T\mu\02\pi}+O(\e e^2 T\mu)+O(e^2 q^2 \ln T) \;\; ,
\eea
where $\s$ is the mass scale introduced by dimensional
regularization. The limit $\e\to0$ is regular so that there arise no
UV singularities proportional to $T$, and the result coincides with
(\ref{p00r}). Actually, in the evaluation of $\d\P$, the use of
dimensional regularization could have been avoided by subtracting
off the unresummed zero-mode contribution, i.e.
\be
  \d \P_{00}(0,q) = 2e^2T \int{d^3k\0 (2\pi)^3}
\lk {1\0k^2+\mu^2}-{1\0k^2} \rk \;\; .
\ee

Compared to the full calculation in the preceding subsection,
the $q$-independence of the next-to-leading order result is
now manifest, which greatly simplifies its evaluation. The
simplifications are less conspicuous in the case of
$\d \P_{ii}(0,q)$, which can be computed in an analogous manner,
readily reproducing (\ref{piir}).

%
%
\let\dq=\thq \renewcommand{\theequation}{5.\dq}    
\setcounter{equation}{0}          

\parag {5. \ Plasma frequency at next-to-leading order }

In this section we shall restrict ourselves to another limiting
case, the one of long wavelengths, $q\to0$, which again simplifies
all calculations considerably.

In this limit $\P\mn$ degenerates and contains only one independent
structure function, because $\P_{00}(Q_0,q\to0)=0$ due to
transversality and $\P_{ij}(Q_0,q\to0)\propto \d_{ij}$ if $\P\mn$ is
to remain regular. Since $\d_{ij}A^{ij}=\d_{ij}B^{ij}$, this entails
that the longitudinal and the transverse polarization function
coincide, $\P_t(Q_0,0)=\P_\ell(Q_0,0)$. In other words, without
a wave vector there is no way to tell longitudinal photonic
quasi-particles from transverse ones.

At $q=0$, the unimproved one-loop result for $T\gg Q_0\sim m$ reads
\be \label{plwlunr}
  \P_{t,\ell}(Q_0,0) = {e^2T^2\09}-{e^2T\012\pi}iQ_0
  - {e^2\024\pi^2}Q_0^2 \ln{\s \0 T} + O(e^2 Q_0^2 T^0)
  = m^2-{e\04\pi}iQ_0m + O(e^2 m^2) \;\; .
\ee
The mass $m$, often identified as {\it the} plasmon mass (although
there is no longer any momentum independent notion of mass)
determines the plasma frequency, below which the medium cannot
sustain free oscillations. At relative order $e$, (\ref{plwlunr})
implies a non-zero damping constant
\be \label{gunr}
 \g = -{1 \0 2m} {\rm Im}\;\P_{t,\ell}(Q_0=m,0)
    = {e^2T\024\pi} = {e\08\pi}m \;\; .
\ee
However, we shall presently show that the subleading term
is not stable under resummation. In particular, the corrected
$\g$ will turn out to vanish at relative order $e$.

\vspace{1cm}
\noindent  
\parbox{16cm}{5.\,A \ RESUMMATION \vspace{.7cm} }
\hfill \vphantom{a} \nopagebreak \indent
Proceeding as in the previous section we find for the resummed
polarization functions
\bea \label{plwlr}
  \P_t(Q_0,0) & = & \P_\ell(Q_0,0) = e^2\sum \lb
  -{4\03}k^2 \D\D^- -{2\D} \rb \nonu\\
  & = & {e^2\03\pi^2}\int_0^\infty k^2 dk {n(\wu{k^2+\mu^2})\0
  \wu{k^2+\mu^2}}\(2+{\mu^2-(Q_0/2)^2\0k^2+\mu^2-(Q_0/2)^2}\) \;\; .
\eea
Because of $\vc q=0$, the angular integration is trivial this time,
and after separating off the constant $T^2$-contribution, the
$O(T)$-part is easily evaluated, yielding
\be \label{plwlr2}
 \P_t(Q_0,0) = \P_\ell(Q_0,0) = {e^2T^2\09}+{e^2T\02\pi}\lb
 -\mu-{4\03Q_0^2}\( [\mu^2-(Q_0/2)^2]^{3\02}-\mu^3 \) \rb \;\; .
\ee

At $Q_0=m$ this determines the correction to the plasma frequency,
\be  \label{5.37}
  m^2+\d m^2 = {e^2T^2\09}\(1-{8\wu2-9\02\pi}e\)\approx
  {e^2T^2\09}(1-0.37 e) \;\; ,
\ee
as well as the $O(e^2T)$-contribution to the plasmon damping in the
long-wavelength limit. However, (\ref{plwlr}) is real at $Q_0=m$,
so that the damping constant vanishes at this order of magnitude.
The bare result (\ref{gunr}) is wrong because by not taking into
account the thermal masses of the scalar particles a plasmon seems
capable of decaying into those. In view of the long story of the
plasmon damping puzzle in hot QCD \cite{samm} let us make the
trivial remark that the manifest gauge independence of the
bare result (\ref{gunr}) did not prevent it from being incomplete.
The nonvanishing result for the resummed QCD damping constant
by the way arises from the possibility of Landau damping, which
is absent for thermal scalars. Thus in scalar QED an imaginary
part to the polarization tensor appears only when pair decay becomes
possible. Indeed, for $Q_0>2\mu$ (\ref{plwlr2}) does become complex,
but $m<2\mu$.

In pure QCD at high temperature the next-to-leading contribution
to the plasma frequency has recently been calculated by one of 
the present authors in Ref.~\cite{HS}, yielding $(\d m^2/m^2)_{\rm
QCD} \approx -0.18 \wu{g^2N}$. Let us see how far this latter
result can be understood by the above result on scalar QED. From
the leading order terms it is clear that $e^2$ corresponds to
$g^2N$, so we might try to apply (\ref{plwlr2}) by inserting the
plasmon mass in place of the thermal mass of the scalars. This
would give $(\d m^2/m^2)\approx -0.028 \wu{g^2N}$, which is over a
factor of 6 short of the actual result. Hence, the correction to
the QCD plasma frequency is much larger than what might be expected
from just the appearance of thermal masses in the loop integrals.

\vspace{1cm}
\noindent  
\parbox{16cm}{5.\,B \ STATIC RESUMMATION \vspace{.7cm} }
\hfill \vphantom{a} \nopagebreak \indent
In the preceding chapter we have seen that the next-to-leading order
results on static Green's functions could be obtained also in a
simplified scheme that resums only static modes. In the
imaginary-time formalism, nonstatic modes are automatically hard so
that their hard-thermal-loop corrections can be treated as
perturbations. This scheme is particularly advantageous in gauge
theories, where it simplifies tremendously the calculation of static
quantities and also of Green's functions with hard external
frequencies. However, dynamical Green's functions with soft external
frequencies are quite intractable because they require analytic
continuation from the imaginary Matsubara frequencies which are
either zero or hard. The separation into static and nonstatic
contributions renders this analytic continuation to general soft
external frequencies rather impossible.

One might perhaps nevertheless expect that also in the nonstatic
situation it is the resummation of internal static modes that gives
the next-to-leading order correction which is nonanalytic in the
coupling constant, and perform the analytic continuation on just
their contribution. This is wrong, as we shall show in the present
example of $\P_{t,l}(Q_0,0)$.

Taking only the $n=0$-contribution of the sum in (\ref{plwlr}) and
continuing at once to soft $Q_0\not=0$ would give
\be \label{plwlstr}
  \d\P_{t,\ell}(Q_0,0) = -{1\03}e^2T\int{d^3k\0(2\pi)^3}\lb
  {4k^2\0 (k^2+\mu^2)(k^2+\mu^2-Q_0^2)}-{6\0k^2+\mu^2} \rb \;\; ,
\ee
where dimensional regularization is understood (see above).
This is readily evaluated, yielding 
\be
  \d\P_{t,\ell}(Q_0,0)\Big|_{\rm static \; contr.}
  = -{e^2T\06\pi}\lb {2\0Q_0}\lk {\mu\0Q_0}
  - \wu{ {\mu^2\0Q_0^2}-1 } \,\rk (Q_0^2-\mu^2)+\mu \rb \;\; .
\ee
It obviously disagrees with (\ref{plwlr2}). It has a wrong analytic
structure in that the imaginary part sets in already at $Q_0\ge\mu$
rather than $2\mu$, and of course does not reproduce the actual real
contribution either. For instance, it would predict the correction
to the plasma frequency squared to be
\be
  m^2+\d m^2\Big|_{\rm static \; contr.} = {e^2T^2\09}
  \( 1-{5\wu5-9\08\pi} e\)\approx {e^2T^2\09}(1-0.09 e) \;\; ,
\ee
which underestimates the next-to-leading order term by more than
a factor of 4.

Evidently, the nonstatic modes may not be neglected and even give
the largest contribution in this example. (In Ref.~\cite{EHKT},
the importance of the nonstatic modes was also noticed in the
case of the gluonic plasmon damping.) The failure of the above
reasoning may be traced to the premature analytic continuation (a
collection of similar pitfalls with analytic continuation at finite
temperature can be found in \cite{Weldonmis}). In our case this
analytic continuation was not possible because the polarization
function in the imaginary-time formalism was given only at one point
in the complex plane of {\it soft} energies, $Q_0=0$, which of
course cannot be continued unambiguously into a function over
nonzero (soft) frequencies.

\vspace{1cm}
\noindent  
\parbox{16cm}{5.\,C \ CLASSICAL PHYSICS  \vspace{.7cm} }
\hfill \vphantom{a} \nopagebreak \indent
We close this chapter by trying to understand the results on the
plasma frequency in the familiar intuitive terms of classical
physics rather than through full-fledged quantum (field) theory.
Looking at the hot scalar plasma as a simple system of particles
interacting through Coulomb forces explains fully the leading order
result $\o^2=e^2T^2/9$, as we shall see, and also part of the
next-to-leading order result.

First, to count the states of scalar particles, imagine a box of
volume $V$ inside the hot plasma and apply periodic boundary
conditions. Then
\be  \label{5N}
  N = {V \0 (2\pi)^3} \int\! d^3k \, n \( \wu {\mu^2+k^2} \)
\ee
is the number of positive scalar particles in $V$ as well as that
of the negative ones. We shall consider again massless particles;
the mass $\mu$ in (\ref{5N}) is to make room for a dynamically
aquired self-energy. Admittedly, the Bose distribution function $n$
stems from quantum physics, but this is the only exterior element
we shall employ.

If a longitudinal electric plane wave $\vc E = E_0 \vc e _1 \sin
(qx-\o t)$ with an infinitesimal amplitude $E_0$ is somehow
activated in the medium, then Maxwell's equations tell us that there
is no magnetic field associated with it. They reduce to
\be  \label{5Max}
   E_0 q \cos (qx-\o t) = \rho \quad , \quad
   j_1 = E_0 \o \cos (qx-\o t) \;\; .
\ee
In the long-wavelength limit $q \to 0$ the charge density
$\rho$ vanishes. For the first component $j_1$ of the current
density we may write $j_1=2e(N/V)v_1$, the factor of 2 coming
from the oppositely charged particles ($-e$) moving with the
opposite velocity $-v_1$. By whatever Newtonian dynamics the Maxwell
equations (\ref{5Max}) are accompanied, it will lead to a factor
$\lambda$ of proportionality between $\6_t \vc v$ and the
force to the particles. Combining this, we obtain the plasma
frequency $\o$:
\be  \label{5om}
 \6_t \vc v = \lambda e \vc E \qquad \Rightarrow \qquad
 \o^2 = \lambda 2 e^2 {N \0 V} \;\; .
\ee
Guessing the inverse mass $\lambda$ to be simply $1/T$, and
determining $N/V$ from (\ref{5N}) as $\approx T^3/8.2$ leads to
$\o^2\approx e^2 T^2 / 4.1$ $-$ which is wrong. The simple
resolution to this puzzle is that one has to take the relativistic
version $\6_t \vc k = e \vc E$ of Newtons equation and
identify the velocity $\vc v$ with a mean value:
\be  \label{5v}
  v_1 = {1 \0 N} {V \0 (2\pi)^3} \int\! d^3k \, f(\vc k , t)
        \, {k_1 \0  \wu {\mu^2+k^2} } \;\; ,
\ee
where $f$ is the distribution function (which in equilibrium is the
Bose function) and $k_1 / \wu {\mu^2+k^2} $ is nothing but the
relativistic velocity-momentum relation. Particles when accelerated
carry their probability with them, so
\be  \label{5fp}
 \6_t f = - \nabla_k f \; \6_t \vc k = - e E_1
 \6_{k_1} n \( \wu {\mu^2 + k^2} \) + O(E^2) \;\; .
\ee
Using this in (\ref{5v}) and (\ref{5om}) we end up with
\be  \label{5kap}
  \lambda = - {1 \0 3} {V \0 N} {1 \0 2 \pi^2}
           \int\! dk \, k^2 \, {k^2 \0 \mu^2 + k^2}
           \, n^\prime \( \wu {\mu^2+k^2} \) \;\; ,
\ee
which when taken at $\mu=0$ gives $\lambda = (V/N) T^2/18$. Thus we
obtain indeed $\o^2 = e^2 T^2 /9$, where one factor 1/3 came from
averaging over the directions of $\vc k$ and the other from the
integration in (\ref{5kap}).

Trying now to go beyond leading order, we could take into account
the thermal mass $\mu=eT/2$ acquired by the scalar particles.
{}From formula (\ref{5kap}) we then obtain for the 'classical'
next-to-leading-order plasma frequency
\be  \label{5class}
  \o^2_{\rm classical} = {e^2 \0 3 \pi^2} \int\! dk \, k^2
   \, { n \( \wu {k^2+\mu^2} \) \0 \wu {k^2+\mu^2} }
   \( 2 + {\mu^2 \0 k^2+\mu^2 } \) \;\; ,
\ee
which is to be compared with the true result (\ref{plwlr}). The only
difference is the missing back-reaction in the form of corrections
involving $Q_0=\o_{\rm classical}$ in the integrand. In (\ref{plwlr})
these terms give rise to an imaginary part when $Q_0>2\mu$, which
obviously corresponds to pair creation. This cannot be captured in
purely classical terms, of course.

Despite the missing terms, (\ref{5class}) is extremely close to the
correct result (\ref{5.37}), to wit
\be
 m^2 + \d m^2 \vert_{\rm classical} 
     = {e^2 T^2 \0 9} \( 1 - {9 \0 8\pi } e \)
     \approx {e^2 T^2 \0 9} \( 1 - 0.36 e \) \;\; .
\ee

%
%
\let\dq=\thq \renewcommand{\theequation}{6.\dq}
\setcounter{equation}{0}   \setcounter{figure}{2}

\parag {6. \ The complete plasmon spectrum at next-to-leading order}

Because of the simplicity of the resummed scalar propagator
determining the two polarisation functions $\P_\ell$ and $\P_t$ up
to and including the next-to-leading order terms, these can in fact
be calculated analytically for the entire $\o$-$q$-plane. In
contrast to the QCD counterparts, the spectral density of the dressed
scalar propagator $\D = 1/(K^2-\mu^2)$ has no cut contribution:
\be  \label{6spectral}
 \D (K) = \int \! dx \, {1 \0 K_0 - x } \, \rho (x,k) \quad ,\quad
 \rho (x,k) = {1\0 2\wsc } \lk \D (x-\wsc ) -\D (x+\wsc ) \rk \;\; ,
\ee
where $\wsc \equiv \wu {\mu^2 + k^2} $. $\P_\ell$ and $\P_t$ as given
in (\ref{2Psubt}), (\ref{2Psubg}) and (\ref{2Psubell}) are determined
by the two sums (and integrals)
\be  \label{6expr}
  \sum \D^- \D  \qquad \mbox{and} \qquad \sum \D^- \D
  \lk k^2- {(\vc k \vc q )^2 \0 q^2} \rk \;\; .
\ee
Here, the upper index on $\D^-$ refers to the shift $K_0 \to Q_0-K_0$
as well as to $\vc k \to \vc q - \vc k$. Hence, the propagator
$\D^- $ has the spectral representation (\ref{6spectral}) with $K_0$
shifted and with the frequency $\wsc _- \equiv \wu {\mu^2 +(\vc k
- \vc q )^2}$ in place of $\wsc$. Equivalently, we may view it as
'another propagator' taken at $Q_0 - K_0$ and with a spectral density
$\rho_-$ as described. At this point we recall two formulae derived
earlier, (6.6) and (6.8) in Ref.~\cite{HS}. They concern the real
and imaginary parts, respectively, of expressions like (\ref{6expr}).
The two are related by the dispersion relation
\be  \label{6disp}
  \Re e \sum \D^- \D f(\vc k ) = \int \! dt \, {1 \0 t-\o} \,
  {1\0 \pi } \lk \Im m \sum \D^- \D f(\vc k ) \rk _{\o =t} \;\; .
\ee
We may therefore concentrate on
\be  \label{6imag}
  \Im m \sum \D^- \D f (\vc k ) = \pi \o T \( {1 \0 2\pi} \)^3
  \int \! d^3k \, f(\vc k )  \int \! dx \, {1 \0 x ( x-\o )} \,
      \rho (x,\vc k ) \,\rho_- (x-\o , \vc k ) \;\; ,
\ee
where on the right-hand-side only the leading temperature-dependent
part resulting from $n(x) \approx T/x$ has been included, assuming
that the integrations are restricted to soft arguments of the Bose
function either automatically or after suitable subtractions.

The details of first evaluating the simpler imaginary part
(\ref{6imag}) and then using (\ref{6disp}) are given in the Appendix
B. The results may be put together as follows. The prefix $\d$ again
indicates that the known leading hard-thermal-loop contributions
have been subtracted. Three regions in the $\o$-$q$-plane are to be
distinguished: $\o^2 < q^2 \,$(region I), $q^2 < \o^2 < 4\mu^2 +
q^2 \,$(region II), and $4\mu^2 + q^2 < \o^2 \,$(region III). Then:
\bea  \label{6res_l}  \hspace{-.3cm}
 \d \P_\ell & = & {e^2 T \0 8 \pi} \; {\o^2 - q^2 \0 q^2}\,\( 4 \mu
 + 2 i {\cal E} - {\o^2 \0 q} \lk {\cal R} + i {\cal J}\rk \) \;\; ,
    \hspace{5.7cm} \\ \label{6res_g}  \hspace{-.3cm}
 \d \P_g & = & {e^2 T \0 8 \pi} \( - 8 \mu + {4\mu^2 + q^2 - \o^2
 \0 q} \lk {\cal R} + i {\cal J} \rk \) \;\; ,
    \hspace{5.7cm} \\  \label{6res_t}  \hspace{-.3cm}
 \d \P_t & = & {1\0 2} \( \d\P_g - \d\P_\ell\)  \nonu \\
  & = & {e^2 T \0 16 \pi} \( - 4\mu {\o^2 + q^2 \0 q^2} - {\o^2 -q^2
  \0 q^2} 2 i {\cal E} + \lk 4\mu^2 + { (\o^2 - q^2 )^2 \0 q^2} \rk
  {1 \0 q} \lk {\cal R} + i {\cal J} \rk \) \;\; , \;\;
\eea
where
\bea  \label{6R}
 {\cal R} & \equiv & \lb \begin{array}{ll}
 \dis\arctan \( {\O + q\0 2 \mu} \) -\arctan \( {\O - q\0 2 \mu} \)
   & \quad \mbox{\frm in I and III}   \\
 \dis 2 \arctan \( {q \0 2\mu + \vert \O \vert } \)
   & \quad \mbox{\frm in II} \lower 14pt\hbox{} \end{array}\right.
   \\  \label{6J}
 {\cal J} & \equiv & \lb \begin{array}{ll}
 \dis\ln \left| {\O q + \o^2 \0 \O q - \o^2} \right|
   & \quad \mbox{\frm in I and III}  \\
 0 & \quad \mbox{\frm in II} \end{array} \right.
   \\  \label{6E}
 {\cal E} & \equiv & \lb \begin{array}{ll}
 \dis\O  & \quad \mbox{\frm in I and III} \\
 \dis i \vert \O \vert & \quad \mbox{\frm in II} \end{array} \right.
\eea
with $\O$, if real, the positive square root of
\be  \label{6Om}
  \O^2 = \o^2 \; {\o^2 - q^2 - 4\mu^2 \0 \o^2 - q^2} \;\; .
\ee
As one learns in Appendix B, the combination ${\cal R} + i {\cal J}$
can be cast into the compact form
\be     \label{6aesthetics}
  {\cal R} + i {\cal J} \, = \, i \, \ln \(
  { 2\mu - i {\cal E} -iq \0 2\mu - i {\cal E} +iq } \) \;\; .
\ee

The static limit of these results obviously reproduces the ones
derived before, (\ref{p00r}) and (\ref{piir}), whereas a bit of
calculation is required to verify that the long-wavelength limit
$q\to 0$ indeed gives $\d\P_\ell(Q_0,0) = \d\P_t(Q_0,0)$ and
coincides with the result derived in (\ref{plwlr}).

Notice that $\d \P_\ell$ and $\d \P_t$ are purely real in region II.
In region I, where the hard-thermal-loop contribution has an
imaginary part corresponding to Landau damping, there is now a
correction term of relative order $e$, whereas the imaginary part
appearing in region III is the leading term resulting from the
possibility of the decay of virtual photonic plasma excitations
into pairs of scalar quasi-particles.

\begin{figure}[t]                                                
\caption[f3]{\label{f3}{\frm The real part of $\d\P_\ell$ in
 units of $e(eT)^2$. $\o$ and $q$ are in units of $eT$.}}        
\end{figure}                                                     

The real part of $\d\P_\ell$ and $\d \P_t$ is displayed in figs. 3
and 4, respectively. For a comparison to the leading-order result
see Fig.~2. In Sect.~5 we have seen that the effect of the
next-to-leading order contributions is to lower the plasma frequency
according to (\ref{5.37}). We shall now study the whole spectrum
of propagating photonic quasi-particles, i.e. $\vc q\not=0$.

\begin{figure}[t]                                                
\caption[f4]{\label{f4}{\frm The real part of $\d\P_t$
             in the same units as used in Fig.~3.}}              
\end{figure}                                                     

In the case of transverse photonic excitations, it turns out that
for increasing $q$ the ratio of the corrected frequency $\o_t(q)$
to the lowest-order one decreases somewhat. For $q\gg eT$ the
effective thermal mass becomes momentum-independent, and the 
dispersion curve goes over into a perfect mass hyperboloid with
asymptotic mass
\be \label{mtasympt}
  m^2_{\infty}+\d m^2_{\infty} = {e^2T^2\0 6} \( 1
  - {3e\0 2\pi} \)  \;\; .
\ee

We now turn to the the more interesting longitudinal plasmons, which
are collective modes without analogues at zero temperature. As we
have seen in chapter 2, their leading-order dispersion curve
approaches the light-cone exponentially with increasing $q$. At the
same time the residue of the corresponding pole in the propagator
vanishes exponentially \cite{Pisres}, quickly rendering them
unimportant for larger values of $q$.

Including the next-to-leading order terms now, we expand everything
around the leading-order result $\o_0(q)^2$ and rewrite the
condition $\o^2 = q^2 + \P_\ell (\o , q )$ for fixed $q$ as
\be \label{6osquare}
    \o^2 = \o_0^2 + \d \P_\ell (\o_0,q) + (\o^2-\o_0^2)
    \; \6_{\o_0^2} \P^{\rm hard}_\ell (\o_0,q) + O(e^2 m^2)
\ee
which gives
\be \label{dom}
  \d \o^2 \;\equiv\; \o^2-\o^2_0 \; = \; { \d\P_\ell(\o_0,q) \0
 1 - \;\6_{\o_0^2} \P^{\rm hard}_\ell (\o_0,q)} + O(e^2 m^2) \;\; .
\ee
In order to work consistently at relative order $e$, we drop all
contributions that are of higher order. With $\P_\ell^{\rm hard}$
given by (\ref{plhtl}), ~(\ref{dom}) can be rewritten as
\be
  \o^2 \; = \; q^2+ \P_\ell^{\rm hard} (\o_0 , q) + {2 \o_0^2 \0
  3m^2 + q^2 - \o_0^2 } \, \d \P_\ell (\o_0 , q ) \;\; .
\ee

The resulting dispersion curves for $e=0.3,\,1,$ and $2$ are shown
in Fig.~5. It is seen that the corrected curves are not just a
slightly down-scaled version of the leading-order result. The
dispersion curves now hit the light-cone at a finite value of $q$,
above which there is no longer any solution corresponding to
propagating plasmons, since for $\o<q$, $\P_\ell$ has a large
imaginary part $\sim e^2T^2$, which prevents the existence of
weakly damped excitations.

\begin{figure}[t]                                                
\caption[f5]{\label{f5}{\frm Leading and next-to-leading order
   dispersion curves of the longitudinal plasmons. The upper curve
   is the leading-order result; below it are the corrected ones for
   $e=0.3,\,1,$ and $2$, respectively. The version on the left
   consistently  discards contributions to $\d\o^2(q)$ that are
   beyond relative order $e$; the one on the right gives the
   location of the poles in the propagator when all
   next-to-leading order contributions to the self-energy
   are kept.}}                                                   
\end{figure}                                                     

The residue of the plasmon pole can be defined in a gauge-independent
manner by projecting the propagator $G\mn(Q)$ onto conserved currents
$J$ with $Q^\mu J_\mu=0$ and restricting to spatially longitudinal
$J_\ell^i = {q^iq^j \0 q^2} J^j$. {} From the component of the
propagator associated with $J_\ell^2$ in
\be
  J^\mu G\mn(Q) J^\nu = -{Q^2\0 Q_0^2} \vc J _\ell^2 \D_\ell
  + \ldots \;\; ,
\ee
we extract the residue as
\be \label{zl}
  Z_\ell = \lim_{Q_0\to \o(q)} {Q^2\0 Q_0^2}
           {Q_0^2 -\o^2(q) \0 Q^2-\P_\ell} \;\; .
\ee
Other definitions, differing in normalization, are possible
\cite{Pisres}, but the one given here avoids unnecessary kinematical
singularities. Abbreviating $\Phi\equiv \o_0^2 \P_l(\o_0,q)/(\o_0^2
-q^2)$, (\ref{zl}) gives
\be
 Z_\ell(q) = Z_\ell^{\rm hard} \lk 1 + Z_\ell^{\rm hard} \(
 \6_{\o_0^2} (\d\Phi)+\d\o^2 \6^2_{\o_0^2} \Phi^{\rm hard}_\ell
 \) \rk \;\; ,
\ee
where $Z_\ell^{\rm hard} = 1-\6_{\o_0^2}\Phi^{\rm hard} \,$.

In Fig.~6 the leading-order result and the corrected one are
compared for the same values of $e$ as in Fig.~5. For both, the
residue decreases rapidly for increasing $q$, with the
one-loop-resummed result being smaller than the leading-order one
for all $q>0$ and arriving at zero at a finite value of $q$.

\begin{figure}[t]                                                
\caption[f6]{\label{f6}{\frm The residues $Z_\ell(q)$ of
                   the plasmon poles  associated with the
                   dispersion curves of Fig.~5.}}                
\end{figure}                                                     

Both results indicate that there is now only a finite range of $q$
and $\o$ in which longitudinal plasmons can exist. However, the very
fact that the corrected result changes qualitatively signals that a
break-down of the {\it resummed} perturbation theory is occurring.
This qualitative change is possible only by $\d\P_\ell/Q^2$ becoming
greater than $\P_\ell^{\rm hard}/Q^2$, although the former is down
by one power of $e$. The reason is that $\P_\ell^{\rm hard}/Q^2$ is
logarithmically divergent for $Q^2\to0$, but $\d\P_\ell/Q^2$
diverges even stronger, like $1/\wu{Q^2}$ (see (\ref{6res_l}) and
(\ref{6E})). Hence, $\wu{Q^2}/q \sim e$ eventually
(over-)compensates for the factor of $e$ in $\d\P_\ell$. In fact,
had we solved $Q^2=\P_\ell^{\rm hard} +\d\P_\ell$ for a small finite
value of $e$ without expanding $\o(q)=\o_0(q)+e\o_1(q)+\ldots$ as we
have done above, the result, shown in the right half of Fig.~5,
would have been that the dispersion curve follows the result
$\o_0(q) + e\o_1(q)$ closely until $Q^2/q^2$ becomes comparable to
$e$, after which it bends back sharply, running down hard by the
light-cone back to $\o=q=0$. If the higher-order corrections to
$\P_\ell/Q^2$ keep being more and more singular for $Q^2\to0$, this
kinky behavior may change from order to order. It is clear that at
any finite order one could trust the result only up to a certain
small distance from the light-cone.

In fact, this break-down of our resummed perturbation theory is
not actually due to higher-order diagrams, but is caused by the
break-down of the high-temperature expansion of $\P\mn$ at $Q^2=0$.
Since $\P_\ell/Q^2\equiv -\P_{00}/q^2$ and the internal lines in
$\P\mn$ have been dressed by their thermal masses, a kinematical
singularity at $Q^2=0$ should not be there at all. Indeed,
as shown in Appendix C, performing the high-temperature expansion
directly at the light-cone $Q^2=0$, i.e. evaluating
$\P\mn(Q_0 = q,q)$, the leading and next-to-leading terms are found
to be
\be \label{pllcres}
  \lim_{Q_0\to q}{\P_\ell^{\rm resum.}\0 Q^2} = {e^2T^2\0 3q^2}
  \lk \ln{2T\0 \mu}+{1\0 2}-\g_E+{\z'(2)\0 \z(2)} \rk
  - {e^2T\mu\0 2\pi q^2}+O(e^2 q^2 T^0)
\ee
with $\g_E$ being Euler's constant and $\z$ the Riemann zeta
function. The first term on the right-hand-side is $\sim\ln(1/e)$,
so the original logarithmic singularity evidently got cut off at
$\wu{Q^2}/(eT)\sim e$ by the resummation of $\mu\sim eT$. The
equation $1=\P_\ell/Q^2$ with $Q^2\to0$ therefore has the solution
\be \label{qcrit}
  q^2_{\rm crit.}/(eT)^2 = {1\0 3} \lk \ln{4\0 e}
  + {1\0 2}-\g_E+{\z'(2)\0\ \z(2)} \rk
  - {e\0 4\pi} = {1\0 3}\ln{2.094\ldots\0 e}-{e\0 4\pi} \;\; .
\ee
Because higher-order diagrams are not singular at $Q^2=0$, either,
the result (\ref{qcrit}) is stable up to and including order $e$. 
The fact that it is non-analytic in $e$ makes it clear that it could
not be obtained by organizing the resummed perturbation theory in
powers of $e$. On the other hand, the strictly perturbative
next-to-leading order result, depicted in the left half of Fig.~5,
becomes inaccurate only close to the light-cone. Indeed, the
``perturbative'' result for $q_{\rm crit.}\approx 0.77 eT$ for
$e=0.3$ is not far off from the one of Eq.~(\ref{qcrit}), which
gives $q_{\rm crit.}\approx 0.79eT$. The alternative result in the
right half of Fig.~5, however, where $\d\o^2$ was not truncated at
relative order $e$, but the entire next-to-leading order result for
$\d\P_\ell$ was kept to define a new full propagator, turns out to
be qualitatively wrong.

The fact that the resummed result for $\P_\ell/Q^2$ is no longer
singular at the light-cone also implies that the residue of the
plasmons does not completely vanish there. This means that there
are longitudinal massless photonic excitations with a fixed value
of $q=\o$. For larger values of $\o$ and $q$, $Q^2$ becomes
negative, i.e. space-like, and Landau damping sets in. At the level
of hard thermal loops, the imaginary part $\sim e^2T^2$ switches on
discontinuously, $\Im m \P_\ell(Q^2)/Q^2 = \th (-Q^2) {3\pi\0 2}
m^2 \o / q^3 $. Because the logarithmic singularity in the hard
thermal loop is removed, this discontinuity is smoothed out. As we
show in Appendix C, for $\o<q$ and $q^2-\o^2\ll e^2q^2$, an extra
factor $\exp(-e\wu{q\08(q-\o)})$ arises --- the imaginary part
sets in with all of its derivatives vanishing. There is now a finite
range in the space-like region with small damping. Consequently, the
plasmons are removed from the spectrum only for
$(q-\o)/(eT)\gsim e^2$.

This phenomenon of a finite range of $q$ and $\o$, where
longitudinal plasmons can exist, was predicted in the case of QCD
in \cite{SU,LS}. The simplicity of scalar electrodynamics has
allowed us to study it in full detail. In the case of
non-ultrarelativistic QED (i.e. with electron mass $m>T$), a finite
range for the longitudinal plasmons has been found already in 1961
by Tsytovich \cite{Tsyt}.

The simple step beyond the resummed perturbation that was necessary
in the above calculation of $q_{\rm crit.}$ might perhaps shed some
light on how similar failures with Braaten-Pisarski resummation
could be overcome. In Eq.~(\ref{pllcres}) we have witnessed the
hard thermal loop itself becoming subject to change under
resummation of the hard thermal loops close to its singularities.
So there are cases where the resummed perturbation theory cannot be
organized in excess powers of $e$. A perhaps similar failure was
encountered in a recent attempt to calculate the production rate of
real non-thermal photons in a QCD plasma by Braaten-Pisarski
resummation \cite{BPS}.

We close this chapter by suggesting that such kinematical
singularities in the Braaten-Pisarski scheme for QCD can be
taken care of by extending resummation to the hard thermal loops
themselves in the same way as we have done above for scalar
electrodynamics. At hard loop momenta it is in fact not necessary
to use the full complicated form of resummed propagators and
vertices in QCD for deriving the modification of the hard thermal
loops near the kinematical singularities they otherwise give rise
to. The latter are caused by the masslessness of the propagators in
the bare hard thermal loop, so it will be the true quasi-particle
poles (rather than cut contributions) that are important. At large
loop momenta $\gg gT$, and only there, these are given by
momentum-independent masses (cp.~Eq.~(\ref{mtasympt})), whereas the
additional collective modes have exponentially vanishing residues at
leading order. Only at relative order $O(g)$ the form of the dressed
propagators and vertices at soft momenta become relevant, whereas
the leading terms are determined by simple massive loop integrals.
The QCD result analogous to the part $\propto T^2$ in
(\ref{pllcres}) has the same form for the bosonic (gluon)
contributions with $e^2\to g^2N$, $\mu\to m_{\infty}$. In the case
of fermionic (quark) contributions a similar formula holds, obtained
by replacing $\g_E\to\g_E-\ln2$ and taking now the asymptotic value
of the thermal quark mass. With $N_f$ flavors, and $N_+ \equiv N
+ N_f/2$, this yields at leading order
\be \label{qcritqcd}
  q^2_{\rm crit.} = 3m^2 \lk {N\0 N_+} \ln {2 \wu 6 \0 g \wu{N_+}}
  + {N_f \0 2N_+} \ln {8\0 g\wu{C_F}} + {1\0 2} -\g_E
  + {\z'(2)\0 \z(2)} \rk \;\; ,
\ee
where $C_F=(N^2-1)/(2N)$. (This result differs somewhat from the one
of Ref.~\cite{SU}, but this seems to be due simply to an incorrect
evaluation of the integrals given therein, with which we otherwise
agree.) We intend to cover the case of QCD more fully in a future
publication. Let us just note here that $q^2_{\rm crit.}$ as given
by (\ref{qcritqcd}) goes down to zero when $g=1.48\ldots$ (for
$N=3$, $N_f=0$) or $g=1.31\ldots$ (for $N=3$, $N_f=2$). Clearly,
next-to-leading results are becoming decisive here to determine the
fate of the QCD plasmons.

%
%
\let\dq=\thq \renewcommand{\theequation}{7.\dq}
\setcounter{equation}{0}    \setcounter{figure}{6}

\parag {7. \  Resummation of scalar damping  }

Up to now we have studied exclusively the effects of the thermal
mass $\mu$ acquired by the scalar fields, i.e. of the leading term
of the self-energy of the scalars. A calculation of the
next-to-leading order terms in this self-energy would already
involve the more complicated thermal photon propagator, making it
necessary to resort to numerical integrations. However, the
limiting case of large scalar momenta turns out to be accessible by
purely analytical means.

For external scalar momentum $Q$ with $Q_0\sim T$, the
next-to-leading order terms can be most easily derived by using the
static ring resummation (see Sect.~4B). For this we only need to
resum the zero-mode propagators
\bea
   S(0,k) = {1 \0 k^2+\mu^2} \quad , \quad
   G(0,k) = { U \circ U - g \0 k^2 } + (\a - 1)
            {\widetilde K \circ \widetilde K \0 k^4 } - 
{U \circ U \0 k^2 + m_{el}^2 } \quad
\eea
of scalar and gauge fields, respectively, with the four-vector 
$\widetilde K \equiv K-(K U)U $ and $m_{el}^2=e^2T^2/3$, see
(\ref{2S}),(\ref{2G}) and (\ref{2AD}).

The correction term to the scalar self-energy, for large
$Q_0, q \sim T$, is then given by
\bea   \label{7delx}
 \d\X & = & {e^2 T \0 (2\pi)^3 } \int\! d^3k \lk  g\omn G\mn (K)
        - { (2Q-K)^\mu \, (2Q-K)^\nu \0 (K-Q)^2} G\mn (K)
          \right. \hspace{2.3cm} \nonu\\ & &  \left. {}
        - { (Q+K)^\mu (Q+K)^\nu \0 K^2 - \mu^2 } G\mn^0 (Q-K)
  \;\;  - \;\; \Bigl(\, \mu,m_{el.}\to0\, \Bigr)
\rk \!\lower 10pt\hbox{$_{K_0 = 0}$} \;\; ,
\eea
where, while the first term clearly corresponds to the tadpole diagram,
the loop contributes twice, once with the hard momentum $Q$ running
through the scalar line (second term) and once with $Q$ in the
photon line (third term). Only the propagator that carries the soft
momentum $K$ requires resummation. On mass-shell, $Q^2=\mu^2$,
the real part of (\ref{7delx}) gives the next-to-leading order
term in the thermal mass of energetic scalar particles, yielding
(after some tedious integrations) 
\be
  \d\mu^2 = -\Re e \, \d\X \Big|_{Q^2=\mu^2}
          = -{e^2T \0 2\pi} \( m_{el.} + {1 \0 2}\mu \)
          = - {1 \0 2\pi } \( {4 \0 \wu 3 } + 1 \) e \mu^2 \;\; .
\ee
The imaginary part, which determines the damping rate $\z$ of
energetic scalar particles defined by
\be
  \Im m\,\d\X \Big|_{Q^2=\mu^2}\equiv 2Q_0 \z \;\; ,
\ee
however leads to an infrared-divergent expression, which is in fact
virtually identical to what is obtained in QED or QCD \cite{BMAR}
\bea \label{imxi}
  \z & = & {e^2T\04\pi} \,\Im m\int_{-1}^1 \! dz
         \int_\lambda^\infty \! dk {k \0 z+k/(2q)-i\e}
  \lk {1\0k^2}-{1\0k^2+m_{el.}^2}-(1-\a){z^2\0k^2} \rk  \nonu\\
  & = & {e^2T\04\pi}\lk \ln{m_{el.}\0\lambda}+O(\lambda^0)\rk \;\; .
\eea
There is a logarithmic singularity caused by the absence of screening
for static transverse electromagnetic fields. In non-Abelian theories
the infrared singularity in (\ref{imxi}) could perhaps be removed by
the appearance of a magnetic screening mass, but in the Abelian case
$\P_{ii}(0,k)\propto k^2$. In Ref.~\cite{LS,LS2,APG} it was suggested
that this infrared divergence is instead cut off by the damping of
the energetic particles itself, entailing $\lambda\sim e^2T$ and
\be \label{scdamp}
  \z = {e^2T\0 4\pi}\ln{1\0 e} \;\; .
\ee
However, this cut-off depends on having the external momentum
slightly off the assumed complex pole \cite{BNN} and this in turn
gives rise to gauge dependences at the order $e^2T\lambda^0$. There
is still an on-going discussion on how to remedy this situation and
to become able to go beyond the leading logarithmic term of
(\ref{scdamp}), see \cite{fermdamp}. In particular it seems likely
that at least in Abelian theories the propagators of moving
quasi-particles do not have simple poles on the unphysical sheet but
rather branch singularities.

At any rate, in order to go beyond the approximations leading to
(\ref{imxi}), one evidently should include self-consistently physics
at scales below $eT$. Taking the result of (\ref{scdamp}) for
granted, one has $\z\sim \mu e \ln(1/e) \gg \d\mu$, so seemingly
damping effects are the next important corrections to be included in
an improved resummation scheme. The consequences of this have been
analysed in Ref.~\cite{LS2} in the case of QED, and recently also in
scalar electrodynamics in Ref.~\cite{MC}. Here we shall follow the
lines of Ref.~\cite{LS2} and correct what has been presented in
Ref.~\cite{MC}.

If one tries to improve the propagator for scalars at large momenta
by including the unexpectedly large damping (\ref{scdamp}), one
finds that in contrast to the thermal mass a small but finite
damping constant may even change the leading hard thermal loop
result for small momenta. It does so, however, by violating gauge
invariance \cite{LS2}. In particular one obtains
\be \label{pooz}
  \P_{00}(Q_0,0) = {2i\z\0Q_0+2i\z}{e^2T^2\03} \;\; ,
\ee
which is inconsistent with transversality of the self-energy.
This seems to have gone unnoticed in a recent paper on the analytic
properties of finite-temperature
scalar electrodynamics Ref.~\cite{Niev}, where it was claimed that
inculsion of damping would restore analyticity of the zero-momentum
limit of thermal Green's functions.

In order to include a finite damping self-consistently one obviously
has to consider vertex corrections. Lacking an effective action,
which would furnish corrected Feynman rules and compensating
(``thermal'') counterterms, one may analyse the possible
contributions from vertex functions through the coupled
Schwinger-Dyson equations they have to satisfy. The
Schwinger-Dyson equation for the photon self-energy is depicted in
Fig. \ref{f7}. It involves the fully dressed scalar propagator as
well as bare and dressed vertices. In Ref.~\cite{LS2,MC} it has been
argued that in the infrared the dominant corrections coming from the
vertices are given by ladder diagrams and that their leading terms
can be obtained by solving the Ward identities of vertex diagrams in
terms of the self-energies and discarding possible transverse
contributions.

  \def\do{{$\!$.}} \def\cic{\circle*{.3}}                        
  \def\vertex{\unitlength.45cm \begin{picture}(3,3)
      \put(0,0){\circle{1.27}} \put(-.1,-.5916){\line(0,1){1.1832}}
      \put(-.3,-.5196){\line(0,1){1.0392}}
      \put(.1,-.5916){\line(0,1){1.1832}}
      \put(.3,-.5196){\line(0,1){1.0392}}
      \put(-.5916,.1){\line(1,0){1.1832}}
      \put(-.5196,.3){\line(1,0){1.0392}}
      \put(-.5916,-.1){\line(1,0){1.1832}}
      \put(-.5196,-.3){\line(1,0){1.0392}} \end{picture} }
\begin{figure}[t] \unitlength.87cm \begin{picture}(17.4,3)
\put(1,1.02)\vertex    \put(5.2,1.02)\vertex  \put(11.8,1.9)\vertex  
\put(13.1,1.02)\vertex \put(17.2,1)\vertex
\put(.46,1){\do} \put(.6,1){\do}
\put(1.42,1){\do} \put(1.56,1){\do} \put(2,.9){$=$} 
\put(2.94,1){\do} \put(3.07,1){\do} \put(3.2,1){\do}
\put(4.2,1){\oval(2,1.8)[l]}   
\put(4.2,1.3){\oval(2,1.17)[tr]} \put(4.2,0.7){\oval(2,1.17)[br]}   
\put(5.64,1){\do} \put(5.777,1){\do} \put(6.4,.9){$+$} 
\multiput(7.44,1)(.14,0){9}{\do}  \put(8,1.03){\do}
\put(8,1.81){\circle{1.6}}        \put(9.1,.9){$+\;2$}
\multiput(10.17,1)(.137,0){7}{\do}
\put(11.200,1.000){\do} \put(11.356,1.020){\do}
\put(11.500,1.080){\do} \put(11.624,1.176){\do}
\put(11.720,1.300){\do} \put(11.780,1.445){\do}
\put(11.800,1.590){\do} \put(11.5,1){\oval(2,1.8)[l]}
\put(12.1,1.3){\oval(2,1.14)[tr]} \put(11.5,0.7){\oval(3.2,1.2)[br]}   
\put(13.537,1){\do}  \put(13.674,1){\do}  \put(14.1,.9){$+$} 
\put(15,1){\do}
\multiput(15,1)(.137,0){15}{\do}  \put(16.2,1){\oval(2,1.8)[l]} 
\put(16.2,1.3){\oval(2,1.2)[tr]} \put(16.2,0.7){\oval(2,1.2)[br]}   
\put(17.64,1){\do}   \put(17.787,1){\do}   
\put(4.2,1.9)\cic   \put(4.2,.1)\cic     \put(8,2.5)\cic 
\put(11.8,.1)\cic   \put(10.76,1.7)\cic  \put(11,1.02)\cic  
\put(12.6,1.84)\cic \put(16.3,1.9)\cic   \put(16.3,.1)\cic 
\put(15.9,1.02)\cic   
\end{picture}    
\caption[f7]{\label{f7}{\frm Schwinger-Dyson equation
    for the photon self-energy. The lines with a black
    bullet are the resummed propagators of figure 1.}}           
\end{figure}
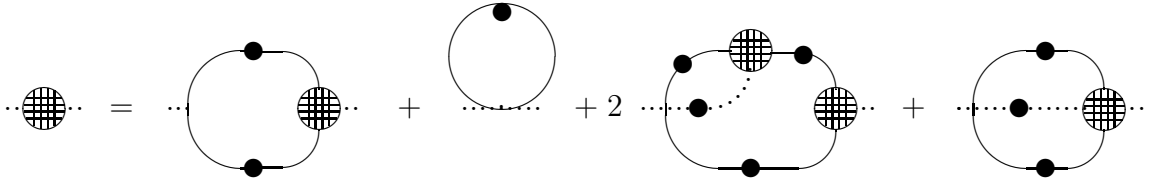                                                     

In order to avoid the intricacies of analytic continuation, in
particular that of vertex diagrams, we shall like
Ref.~\cite{LS2,MC} work in the real-time (Schwinger-Keldysh)
formalism \cite{SchwK}. In this formalism perturbation theory is
based on a time contour that runs along the real axis and back,
which leads to a doubling of the fields corresponding to the part of
the contour they are living on, and to a contour-ordered propagator
that is a matrix with respect to this bisection. If $a,b=1,2$
denotes the field type, a scalar propagator is now given by the
matrix
\be
  -i\D(x-y) = \< T_c\phi_a(x)\phi_b(y) \> =
  \(\begin{array}{cc} \< T\phi(x)\phi(y)\> & \< \phi(y)\phi(x)
  \> \\   \<\phi(x)\phi(y)\> & \< \bar T\phi(x)\phi(y) \>
  \end{array}\)   \;\; ,
\ee
where the contour-ordering $T_c$ is broken up into time-ordering
$T$ for type-1 fields and anti-time-ordering $\bar T$ for type-2.
Mixed propagators correspond always to a fixed order of the
operators because type-2 fields are always further down the
contour than type-1.

Analytic continuation from the imaginary-time formalism eventually
leads to a collection of Green's functions which depend on the
prescriptions chosen for each external line. In the real-time
formalism, these are given by the various linear combinations of the
components of the corresponding matrix quantities \cite{SchwK,RA}.
The retarded and advanced propagators are given by
\be
 \D_R(P) = \D_{11}(P)-\D_{12}(P) \;\; , \quad \D_A(P)
         = \D_{11}(P)-\D_{21}(P) \;\; .
\ee
The symmetric combination, defined by
\be
  \D_P(P) = \D_{11}(P)+\D_{22}(P)\equiv \D_{12}(P)+\D_{21}(P) \;\; ,
\ee
is, in thermal equilibrium, related to the former,
\be
  \D_P(P) = (1+2n(P_0))(\D_R(P)-\D_A(P)) \;\; .
\ee
Whereas at tree level, vertices are exclusively connecting either
type-1 or type-2 fields, there is a large variety of full vertex
functions which differ in their analytic properties. In the first
diagram on the right-hand-side of Fig. \ref{f7} we have a full
vertex function connecting an incoming scalar, a photon, and an
outgoing scalar. Defining retarded vertices with respect to these
three lines in turn we have
\be
  \G_{R(i)} = \sum_{a,b}\G_{1ab} \;\; , \quad
  \G_{R(\g)} = \sum_{a,b}\G_{a1b} \;\; , \quad
  \G_{R(o)} = \sum_{a,b}\G_{ab1} \;\; ,
\ee
where the indices are in the order of incoming scalar (i), photon,
and outgoing scalar (o). There are various symmetric combinations,
of which we shall need only
\be
  \G_P = \sum_{a,b}\G_{aba}
\ee
which is the one that singles out the photon line.

With these definitions the retarded piece of the first diagram on
the right-hand-side of Fig. \ref{f7} is given by
\bea \label{pir1}
 \P\mn^{(1)R}(Q) & \equiv &
 \P\mn^{(1)11}(Q) + \P\mn^{(1)12}(Q) \nonu\\
 & = & {e\0 2} \int{d^4K\0(2\pi)^4}(2K+Q)_\mu \Bigl\{
 \D_R(Q+K)\G^{R(i)}_\nu [\D_P(K)+\D_R(K)] \nonu\\
 & & {} \qquad + [\D_P(Q+K)+\D_A(Q+K)]
     \G^{R(o)}_\nu \D_A(K) \nonu\\
 & & {} \qquad + \D_R(Q+K)\G^P_\nu \D_A(K) \Bigr\} \;\; .
\eea

The second diagram involves only the bare 4-vertex and therefore
differs from the usual tadpole (or sea-gull) diagram only in that
the full propagator has to be taken. In terms of the real-time
quantities it is simply given by
\be \label{pir2}
  \P\mn^{(2)R}(Q) = ie^2g\mn
  \int{d^4K\0(2\pi)^4}[\D_P(K) + \D_R(K)+\D_A(K)] \;\; .
\ee
The other diagrams are of higher (explicit) power in $e$ and we
shall not need them in the following discussion.

We wish to consider a resummation of both thermal masses and the
``anomalously'' \cite{LS2} large damping of the scalars. We
therefore begin by adding the damping to the retarded and advanced
scalar self-energies according to
\be
  -\X_R(P) = \mu^2-2iP_0\z \;\; , \quad
  -\X_A(P) = \mu^2+2iP_0\z = -\X_R^* = -\X_R(-P) \;\; .
\ee
The momentum dependence of these additions spoil the Abelian Ward
identities and this leads to a violation, proportional to $\z$, of
the generally valid transversality of the photon self-energy already
at the order of $e^2T^2$, i.e.~at the level of hard thermal loops
\cite{LS2}. In Ref.~\cite{LS2} it has been shown however that
higher-order vertex diagrams start to contribute because with the
above modification of the propagators there are now contributions
proportional to $\z$ from would-be pinch singularities which are now
cut off by the small but finite damping. It turned out that the
dominant vertex contributions just correspond to solving the Ward
identities according to
\bea \label{wardis}
  \G_\mu^{R(i)}(K+Q,K) & = & -\(\G_\mu^{R(o)}(K+Q,K)\)^* \nonu\\
  & = & -ie(2K+Q)_\mu\(1+{1\02 KQ + Q^2}
  \lk \X_R(K+Q)-\X_R(K) \rk\) \nonu\\
  & = & -ie(2K+Q)_\mu \lk 1+{2i\z Q_0\0 2KQ + Q^2} \rk \;\; , \\
  \G_\mu^P(K+Q,K) & = & {} -ie(2K+Q)_\mu{1\0 2KQ + Q^2}
  \lk \X_P(K+Q)-\X_P(K) \rk \nonu\\
  & = & {-ie(2K+Q)_\mu\0 2KQ + Q^2}(4i\z)\biggl[ Q_0 \nonu\\
  & & {} \qquad
  + 2(K_0+Q_0)n(K_0+Q_0)-2K_0 n(K_0) \biggr] \;\; .
\eea
(In the case of QED \cite{LS2} only $\G^P$ needed correction because
there the damping of the electrons arises from a constant
contribution to the self-energy.)

Inserting the corrected vertices into (\ref{pir1}) and keeping only
terms involving the distribution function $n$, one finds for the
dressed one-loop contribution to the retarded photon self-energy
\bea \label{pir}
  \P\mn^R(Q) & = & -2ie^2
  \int{d^4K\0(2\pi)^4}(2K+Q)_\mu(2K+Q)_\nu \, \biggl\{ \nonu\\
  & & \D^R(K+Q)\lk 1 + {2i\z Q_0\0 2KQ + Q^2} \rk n(K_0)
  \(\D^R(K) - \D^A(K)\) \nonu\\
  & & {} + 2i\z {(K_0+Q_0)n(K_0+Q_0)-K_0n(K_0)\0 2KQ + Q^2}
  \D^R(K+Q)\D^A(K) \biggr\} \nonu\\
  & & {} +2ie^2g\mn
  \int{d^4K\0 (2\pi)^4}n(K_0)\(\D^R(K) - \D^A(K)\) \;\; .
\eea

By a change of variables,
this can be brought into a form that can be more easily compared
with the corresponding expressions given in Ref.~\cite{MC},
\bea \label{piru}
  \P\mn^R(Q) & = & 2ie^2 \int{d^4K\0 (2\pi)^4}(2K+Q)_\mu (2K+Q)_\nu
                 n(K_0) \, \biggl\{ \nonu\\
  & & \D^R(K+Q)\D^R(K){4i\z K_0\0 2KQ + Q^2} \nonu\\
  & & {} - \D^R(K+Q)\(\D^R(K) - \D^A(K)\) \lk 1+2i\z{2K_0
         + Q_0\0 2KQ + Q^2} \rk  \biggr\} \nonu\\
  & & {} + 2ie^2g\mn
  \int{d^4K\0 (2\pi)^4} n(K_0) \( \D^R(K) - \D^A(K) \) \;\; .
\eea
again up to temperature-independent contributions. This does not
completely agree with Eq.~(25) of \cite{MC}: there the first term
in the first integrand was omitted, and in the second integral a
dressed 4-vertex was used, contrary to what the Schwinger-Dyson
equation prescribes. Moreover, it was assumed in Ref.~\cite{MC}
that the dominant contributions from the vertices would be the
correction terms, because they involve a factor $1/(2KQ+Q^2)$,
so the tree-level vertices were deleted. Further, $(2K+Q)_\mu/(2KQ
+Q^2)$ was simplified to $K_\mu/KQ$. Neither of these truncations
is justified, however, as a complete calculation readily shows. The
part of (\ref{pir1}) which involves only the tree vertices does
produce the same leading temperature results as without resummation
of the scalar damping plus terms proportional to $\z$, and the
vertex corrections that are proportional to $\z$ stay so when the
integrals are evaluated. The replacement $(2K+Q)_\mu/(2KQ+Q^2) \to
K_\mu/KQ$ on the other hand changes the contributions coming from
the vertex corrections, but even after discarding the pure
tree-level contributions the leading temperature contributions are
not restored. For instance, the Debye mass $\P_{00}$ comes out minus
half the correct value. In Ref.~\cite{MC}, the missing factor of 2
was compensated by taking the first diagram in the right-hand-side
of Fig. \ref{f7} twice; the minus sign seems to have been simply
lost.

Another thing that goes wrong with the expressions put forward in 
Ref.~\cite{MC} is in fact gauge invariance. As a special case of the
transversality of the photon self-energy one has to have $\P_{00}
(Q_0,0)\equiv0$. The expressions given in Ref.~\cite{MC} violate
this requirement already in the leading terms of the high-temperature
expansion, both with and without the unjustified simplifications
performed therein: $\P_{00}(Q_0,0)$ in each case turns out to be
proportional to $ie^2T^2\z/Q_0$.

Coming back to the complete results (\ref{pir},\ref{piru}), they
can be evaluated by closing the contour in the $K_0$-integration.
In the upper half plane, there are poles from the advanced Green
fucntion $G^A(K)$ and an infinite set of poles from the Bose
distribution function $n$. There are also potential poles from the
denominators of the vertex corrections, which are ambiguous because
the Ward identities do not determine the prescription for these.
We shall therefore consider only such cases where these denominators
cannot give rise to poles on algebraic grounds. This holds generally
for $Q_0=0$ and with $Q_0\not=0$ at least for $\P_{00}(Q_0,0)$,
which is of special interest as a non-zero result indicates a loss
of gauge invariance.

In the form (\ref{piru}) there are nonvanishing contributions from
the poles of the Bose distribution function. For example, with
$\vc q=0$, $Q_0\ll T$, they contribute at leading order in $T$
\bea
  & & {} -16e^2 T\z\sum_{n=1}^\infty
  \int{d^{(3-\e)}k\0(2\pi)^{(3-\e)}}
  {(2\pi inT\d_\mu^0+k_i \d^i_\mu)(2\pi inT\d^0_\nu+k_j\d^j_\nu)\0
  iQ_0((2\pi nT)^2+k^2)^2} \nonu\\
  & = & {ie^2\z T^2\03Q_0}\(\d_\mu^0\d_\nu^0
  + \d_{ij}\d^i_\mu\d^j_\nu\) + O(\e) \;\; ,
\eea
which is due to the first term in the first integral of (\ref{piru}).
The omission of this contribution is what led to the (unnoticed)
loss of gauge invariance in Ref.~\cite{MC}. In the seemingly more
complicated form (\ref{pir}), however, it turns out that in all the
considered cases the contributions from the poles of the distribution
functions which would be either proportional to $T^2$ or $T$ exactly
cancel upon integration over spatial momenta; only the poles of
$G^A(K)$ contribute by closing the contour around the upper half
plane.

At leading order $T^2$, the correction terms to the vertices in
Eq.~(\ref{pir}) are found to exactly cancel the result (\ref{pooz})
which indicated that dressing of only the propagators violates gauge
invariance. However, contrary to what has been claimed both in
Ref.~\cite{LS2} and \cite{MC}, the one-loop resummed expressions are
not completely independent of $\z$. There are still contributions
$\propto T$ in (\ref{pir}) that are dependent on $\z$. Checking
gauge invariance first, we find for the leading term that violates
transversality of the photon self-energy
\be \label{pooz3}
  \P_{00}(Q_0,0) = -{ie^2\z^3T\0\pi\mu(Q_0+2i\z)} \;\; .
\ee
For $Q_0\lsim\z$, this is of order $e^5\ln^3(1/e)T^2$; for larger
$Q_0$ it is of even higher order. (A similar result is obtained
also in the case of QED.)

Inspecting now potential contributions to $\P\mn(0,q)$, we find that
the only new corrections are of precisely the order where gauge
invariance is being lost,
\bea \label{pz3}
  \P_{00}(0,q\to0) & = & {e^2T^2\03}-{e^2T\mu\02\pi}
                     - {e^2T\z^2\0 4\pi\mu} \;\; , \\
  \P_{ii}(0,q\to0) & = & {3e^2T\z^2\08\pi\mu} \;\; .
\eea
The latter result, which if correct would correspond to a magnetic
screening mass, is by the way just another expression of the fact
that gauge invariance has been lost --- in scalar electrodynamics
one can easily prove the absence of a magnetic screening mass to
all orders of perturbation theory \cite{Fradkin,KaKli}.

The form of the above results in fact shows why the resummation of
scalar damping is bound to fail eventually. The contributions 
involving $\z$ are obviously dominated by the infrared region of
integration, which is effectively cut off by the thermal mass $\mu$.
The damping of the scalar fields on the other hand has its simple
constant form certainly only for momenta $\gg eT$, so precisely
where interesting effects might show up, this resummation is clearly
insufficient.

In view of recent attempts to go beyond the Braaten-Pisarski
resummation scheme by a simple resummation of damping effects
\cite{AS}, we close by emphasizing again that a resummation of
damping only in the propagators proves to be incorrect and
insufficient already in the Abelian case. Vertex corrections are of
vital importance to keep gauge invariance, and they tend to just
undo the resummation of damping in the propagators. Moreover, in
the infrared regime where a resummation of damping might eventually
become important, the full-fledged self-energy and vertex functions
will be relevant, which can hardly be approximated by a constant
damping term.

%
%
\let\dq=\thq \renewcommand{\theequation}{7.\dq}
\setcounter{equation}{0}

\parag {8. \  Summary and conclusions  }

The simplicity of hot scalar electrodynamics has allowed us to
determine the complete next-to-leading order corrections to the
spectrum of photonic quasi-particles. For the most part, this was a
straightforward application of the resummation scheme of Braaten
and Pisarski developed for QCD, with the bonus of being able to do
all calculations analytically, so that we did not depend on
numerical approximations. As we have shown in detail in the
Appendix, these calculations can be most efficiently done by
evaluating first the simpler imaginary parts and exploiting
dispersion relations to obtain the complete expressions.

Against these results we tested the simplified resummation scheme
of Ref.~\cite{AE}, which is a systematic extension of the old
ring resummation prescription of dressing zero modes only, 
and found that it works only for static quantities (for which it has
been put forward originally), whereas non-static Green's functions
turn out to receive next-to-leading-order corrections also from
non-static modes. In the case of static Green's functions, on the
other hand, the resummation scheme of Ref.~\cite{AE} constitutes a
computational simplification, which is of course more pronounced in
non-Abelian applications.

The next-to-leading order results for the plasmon spectrum
constituted mostly small corrections to the ones derived from a
bare one-loop calculation, but on two occasions they led to a
qualitative change.

First, the bare one-loop result gave a nonvanishing result for
plasmon damping of the order of $e^2T\sim em$. Like in QCD,
this result turned out to be inaccurate, although unlike the
bare one-loop gluonic plasmon damping, it was positive and
gauge-independent. The resummed result revealed that the photonic
plasmon damping is in fact zero at relative order $e$, because the
thermal scalars are too massive to be produced by pair creation and
also because at this order there is no Landau damping contribution
from scalar quasi-particles. The obvious moral is that
gauge-independence is only a necessary criterion for a correct
result, not a sufficient one. 

Second, the next-to-leading order results for the longitudinal
plasmons turned out to break down very close to the light-cone.
At leading order, the longitudinal plasmon branch of the dispersion
curves approaches the light-cone exponentially. Very close to
the light-cone, the next-to-leading order corrections derived
strictly along the lines of the Braaten-Pisarski resummation scheme
begin to dominate over the leading-order contributions, so
this scheme ceases to be actually perturbative. The reason for
this is a singularity, at the light-cone, in the hard thermal loops
that are being resummed. A more persistent resummation removes this
singularity and allowed us to calculate the true next-to-leading
order dispersion curve, which hits the light-cone at a finite value
of momentum $\sim eT\ln^{1/2}(1/e)$, and continues even a short
distance below the light-cone before Landau damping prohibits the
existence of weakly damped excitations. We have argued that the
same phenomenon occurs in QCD, as has been done before in
Ref.~\cite{SU}, whose corresponding result we believe to have
corrected. The encountered break-down of the canonical
Braaten-Pisarski scheme and its redress could be relevant also with
respect to another recently observed break-down of this scheme in
Ref.~\cite{BPS}, which also is due to an irregular behavior of the
hard thermal loops at the light-cone.

Finally, we have investigated another modification of the
Braaten-Pisarski scheme, which attempts the next step by resumming
also damping contributions from internal lines. In accord with
Ref.~\cite{LS2}, where similar issues have been studied in the case
of fermionic QED, we found that there are no contributions to the
photonic polarization tensor from such a further resummation, if and
only if the vertices are corrected in addition to the propagators.
We disagree with the findings of Ref.~\cite{MC}, however, where this
sort of resummation was examined for scalar electrodynamics. In
contrast to Ref.~\cite{MC}, we found that the corrections to the
vertices do not dominate over tree contributions, an assumption
which led Ref.~\cite{MC} to postulate a new diagrammatic framework.
Instead, both contribute at the same order of magnitude and give
correct results with the conventional set of diagrams as prescribed
by the Schwinger-Dyson equations. Further, we observed that the
assumption of a constant damping term can be sufficient only for
leading-order results, and that it leads to a violation of gauge
invariance at higher orders even with vertices corrected such that
they satisfy the Ward identities. However, without correcting the
vertices, gauge invariance is lost already at lowest order. This
seems to have gone unnoticed in a recent paper on analyticity
properties of scalar electrodynamics, Ref.~\cite{Niev}. It also
casts doubt on recent attempts to improve on perturbative results by
an ad hoc resummation of damping contributions in QCD \cite{AS} only
in propagators.
 
{\bf Acknowledgements}

We are grateful to R. Baier for many stimulating und helpful
discussions.

%
%
\let\dq=\thq  \renewcommand{\theequation}{A.\dq}    
\setcounter{equation}{0}

\parag {Appendix \ A}
  
In this Appendix, which supplements Sect.~3, we briefly go through
the construction of the effective action summarizing the hard
thermal loops along the lines of Refs.~\cite{eff,rel,BPeff}
restricted to the case of scalar electrodynamics. As was first
pointed out in \cite{eff}, this effective action is completely
determined by the bilinear part of the generating functional
$S_\eff^{(2)}$ that is proportional to $e^2T^2$, i.e.~by the HTL
self-energies.

To construct $S_\eff^{(2)}$, or rather its nontrivial part
$\d S^{(2)}$, we sandwich $-\mu^2$ with scalar fields and ${1\0 2}
\P\omn$ with photon fields. With the leading term of (\ref{2Pi}),
if rewritten in terms of $Y$, \cite{FT} one obtains
\be   \label{effG}
  \d S^{(2)} = - \mu^2 \int^\beta \varphi^\ast \varphi + {3\0 2} m^2
  \int^\beta \ell^{\prime\prime} \quad , \quad \ell^{\prime \prime}
  = \int_\O \( A_0 A_0 + ( \6_0 AY ) {1\0 d} AY \)
\ee
\be  \label{AY}  
  \mbox{with} \qquad AY = A_0 - \vc e \vc A \qquad \mbox{and} \qquad
  d \equiv Y_\mu \6_\mu = \6_0 + \vc e \nabla  \;\; .
\ee
Now consider the gauge variations $\d_\o A_0 = \6_0 \o$ \,
and \, $\d_\o AY = d \o \,$. This gives
\be \label{effdo}
  \d_\o \int^\beta \ell^{\prime\prime} = \int^\beta \int_\O
  \( 2 A_0 \6_0 \o + (\6_0 d \o ) {1\0 d}
  AY + \o \6_0 AY \) = 0 \;\; ,
\ee
i.e.~gauge invariance of $S_\eff^{(2)}$. The vanishing of
(\ref{effdo}) is due to $\int_\O \vc e = 0$ in the first term (which
allows us to rewrite it as $2 AY \6_0 \o$) and partial integrations
in the two others. Therefore, gauge invariance does not require the
existence of HTL vertices. Moreover, since all HTL's are uniquely
determined by $S_\eff^{(2)}$ \cite{eff}, this indeed proves the
absence of HTL vertices (cf. Sect. 3).

While the generating functional thus obtained, $S_\eff =
S_\eff^{(2)}$, is gauge invariant, the corresponding Lagrangian
density is not. A manifestly gauge invariant (mgi) density
$\ell^\prime$ can be obtained by adding to $\ell^{\prime\prime}$ a
suitable total derivative $\6_\mu J^\mu$. It is not hard to
construct this current as $J^\mu = -\int_\O Y^\mu A_0 {1\0 d} AY$.
Thus:
\be   \label{effv}
  \ell^\prime = \ell^{\prime\prime} + \6_\mu J^\mu = \int_\O
  \( \6_0 AY - d A_0 \) {1\0 d} AY \;\; .
\ee
(\ref{effv}) is easily checked to be mgi indeed: $\d_\o \ell^\prime
= 0$. But in this check $\int_\O \vc e = 0$ has to be exploited, and
in fact one can do better. The density $\ell^\prime$ can be expressed
entirely by mgi objects, which are the fields $\vc E = -\nabla A_0 -
\6_0 \vc A$ and $\vc B = \nabla \times \vc A $ or combinations
of them as e.g.
\be \label{efff}
  f_\mu \equiv Y^\nu F_{\nu \mu} = d A_\mu - \6_\mu AY
   \;\; , \qquad \mbox{i.e.}
\ee
\be  \label{eff0}
   f_0 = - \vc e \vc E \qquad \mbox{and} \qquad 
   \lb f^k \rb = \vc f = -\vc E - \vc e \times \vc B \;\; .
\ee   

The mgi action in the QCD case was constructed independently by
Frenkel and Taylor \cite{rel} and by Braaten and Pisarski
\cite{BPeff}, where the latter just guessed the only possible form
in accord with known facts and principles. Ref. \cite{rel} also
gives a proof of its equivalence with the non-mgi version. Here we
read off the mgi action for scalar electrodynamics
from eq. (20) of \cite{BPeff}
by taking its abelian equivalent:
\be \label{effvp}
  \ell = - {1\0 2} \int _\O f_\mu {1\0 d^2} f^\mu \;\; .
\ee
Note that the trace over generators has turned into a factor $1/2$.
The two versions $\ell^\prime$ and $\ell$, both mgi, could differ
by a total derivative which is mgi too. However, there is no such
difference, as we show next. This is in fact already covered by the
proof for the non-Abelian case \cite{rel}, but in the much simpler
Abelian case, we can do a straightforward calculation which starts
with '$\ell^\prime =$' and results in '$= \ell \, $'.

In a first step we try to eliminate the potentials from the angular
integral $\ell^\prime$, (\ref{effv}), in favour of the fields
$\vc E$, $\vc B$. This can be done by using the first two of the
following five identities:
\be  \label{A1}  
   {\vc e \0 d} = X \nabla \quad , \quad {\vc e \circ \vc e \0 d}
   = - X \6_0 \( 1 + {3 \0 2} Z \) - {1 \0 2 d} Z \;\; ,
\ee
\be  \label{A2}  
   {\vc e \0 d^2} = {1 \0 d} X \nabla \quad , \quad {1 \0 d^2}
   = {1 \0 \6 ^2} \;\; ,
\ee
\be  \label{A3}  
  {\vc e \circ \vc e \0 d^2} = X \( 2+3 Z\)+{1\0 d^2} \( 1+Z\)\;\; ,
\ee
where
\be  \label{A4}
  X \equiv {1\0 \D } \( 1-{\6_0 \0 d } \) \qquad \mbox{and}
  \qquad Z \equiv {1\0 \D } \nabla \circ \nabla - 1 = {1\0 \D }
  \nabla \times \( \nabla \times \ldots \) \;\; .
\ee
(\ref{A1}) to (\ref{A3}) are replacements allowed under the angular
avergage $\int_\O$ if no further dependence on $\vc e$ occurs. Using
(\ref{A1}) and, in the last step, $Z\vc A = (1/\D ) \nabla \times
\vc B $ one obtains
\bea
 \ell^\prime & = & \int_\O \( \vc E {\vc e \0 d } A_0 - \vc E {\vc e
 \circ \vc e \0 d} \vc A \) \nonu \\
 & = & \int_\O \( \vc E X \nabla A_0 + \vc E X \6_0 \( 1 +
   {3 \0 2} Z \) \vc A + \vc E {1 \0 2d} Z \vc A \)  \nonu \\
 \label{A5} 
 & = & \int_\O \vc E \, \( - X \vc E + \( {3\0 2} X \6_0 +
   {1\0 2d} \) {1\0 \D } \nabla \times \vc B \) \;\; .
\eea  
We get rid of $\6_0$ through the homogeneous Maxwell equation
$\6_0 \vc B = - \nabla \times \vc E $, i.e. $\6_0
(1/\D )\nabla \times \vc B = -  Z \vc E $. Now (\ref{A3}), used in
backward direction, and (\ref{A4}) lead to
\be  \label{A6}
 \ell^\prime = {1\0 2} \int_\O \( \vc E {1\0 d^2} \vc E -\vc e \vc E
 {1\0 d^2} \vc e \vc E + \vc E {\vc e \0 d^2} \times \vc B \) \;\; .
\ee
All three terms are contained also in the desired result $\ell$. But
two terms are missing. So, in the last step, we exploit that
\be  \label{A7}
  \( \vc e \times \vc B \) {1 \0 d^2 } \( \vc e \times \vc B \)
  + \( \vc e \times \vc B \) {1 \0 d^2 } \vc E = 0  \;\; .
\ee
Adding this under $\int_\O$ in (\ref{A6}) the calculation ends up
with
\be  \label{A8}
 \ell^\prime \; = \; - {1 \0 2} \int_\O \lk \vc e \vc E {1 \0 d^2 }
 \vc e \vc E - \( \vc E + \vc e \times \vc B \) {1\0 d^2 } \( \vc E
 + \vc e \times \vc B \) \rk \; = \; \ell \;\; ,
\ee     
see (\ref{effvp}) with (\ref{eff0}).

To verify (\ref{A7}), we use (\ref{A2}) and (\ref{A3}), note that
$Z \vc B = - \vc B $ and eliminate $\vc E $ by means of the
homogeneous Maxwell equation:
\bea   \label{A9}
 \vc B {1\0 d^2 } \vc B - \vc B \vc e {1\0 d^2 } \vc e \vc B - \vc B
 {\vc e \0 d^2 } \times \vc E & = & \vc B {1\0 d^2 } \vc B + \vc B
 X \vc B - \vc B X {1 \0 d } \nabla \times \vc E  \nonu \\
 & = & \vc B \, {1 \0 \D } \( 1 -{\6 ^2 \0 d^2 } \) \vc B
       \, \, = \, 0 \;\; ,
\eea
where (\ref{A2}) explains the last step. (\ref{effG}) with $\ell^{
\prime\prime} \to \ell^\prime = \ell$ and (\ref{effvp}) constitute
the result (\ref{3eff}) given in the main text.

%
%
\let\dq=\thq  \renewcommand{\theequation}{B.\dq}
\setcounter{equation}{0}

\parag {Appendix \ B}

Here the next-to-leading order contributions to the polarization
functions $\P_\ell$ and $\P_t$ 
as given by Eqs.~(\ref{2Psubel}) and (\ref{2Psubt})
are evaluated for all $\o$ and $q$.
The strategy for doing this was already outlined at the beginning
of section 6. We start with the imaginary parts of the two sums
(\ref{6expr}) and apply formula (\ref{6imag}). For convenience we
write
\be  \label{B1}
   \Sigma_i \equiv \sum \D^- \D \; f_i \quad (i=1,2)
   \qquad \mbox{with} \qquad f_1 = 1 \quad \mbox{and} \quad
   f_2 = k^2 - { (\vc k \vc q )^2 \0 q^2} \;\; .
\ee
The two integrals in (\ref{6imag}) allow for the substitutions
$x \rightarrow \o - x$ and/or $\vc k \to \vc q - \vc k $. Doing
both transformations the integrand remains unchanged, since the
$f_i$ are invariants under this transformation and the density
$\rho$ is an odd function of $x$. Thus, the prefactor $\o$ may be
taken inside the integral as $\o = x + (\o -x) \to 2x\,$:
\be    \label{B2}
 \Im m \Sigma_i =  2 \pi T \( {1\0 2\pi} \) ^3 \!
    \int \! d^3k \, f_i \int \! dx \, {1 \0 x-\o}\,
    \rho_- (x,\vc k ) \, \rho (x-\o ,\vc k ) \;\; .
\ee
We insert the spectral densities of (\ref{6spectral}). After some
trivial manipulations with the delta functions (with the aim to
remove $\wsc_- $ from the prefactors) we obtain
\bea   \label{B3}
  \Im m \Sigma_i & = & {T \0 4\pi} \int_0^\infty \!\! dk \, k^2
    \int_{-1}^1 \!\! du \; f_i \, {1 \0 \wsc^2} \,\sgn (\o - \wsc )
        \,\d \( 2kqu + q^2 - \o^2 + 2 \o \wsc \)
        \; - \; (\o \to -\o ) \nonu \\
      & = & {T \0 8\pi q} \int_0^\infty \!\! dk \,
          {k \0 \wsc^2 } \, f_i \, \sgn ( \o - \wsc ) \; \Theta
        \; - \; (\o \to -\o ) \;\; ,
\eea
where, while $f_1$ will always equal 1, the function
$f_2$ has turned into
\be  \label{B4}
  f_2 =  k^2 - {1 \0 4 q^2} \,
      \lk q^2 - \o^2 + 2 \o \wsc \;\rk^2 \;\; .
\ee
$\Theta$ stands for the step function
$ \Theta = \th \( 2kq- \vert \o^2 - 2\o \wsc
          - q^2 \vert \, \)$.
Next we change to the integration variable $x=\wu {\mu^2 + k^2}
= \wsc $ and manipulate the step function:
\be   \label{B5}
 \Theta = \th \, \( \lk q^2 - \o^2 \, \rk \lk \, \(
  x - {\o \0 2} \) ^2 - {q^2 \0 4 \o^2 } \O^2 \, \rk\, \)\;\; ,
\ee
where
$\O^2 \equiv \o^2 (\o^2 - q^2 - 4\mu^2)/(\o^2 - q^2 )$
as given in (\ref{6Om}) in the main text. Obviously (and
notwithstanding the remaining integration), there is no imaginary
part in region II, i.e. in $q^2 < \o^2 < 4\mu^2 + q^2$. For a
convenient formulation in the other two regions we introduce
\be   \label{B6}
   x_1 = {\o \0 2 } - {q \0 2 \o } \, \O
   \qquad \mbox{and} \qquad
   x_2 = {\o \0 2 } + {q \0 2 \o } \, \O \quad , \quad
   \nonu
\ee
so that
\be  \label{B7}
 \Theta = \lb \matrix{
       \th \, \( \, [\, x - x_1 \, ]\, [\, x_2 -x\, ] \, \)
                & \quad \mbox{\frm in III } \;\, \cr
           0    & \quad \mbox{\frm in II  } \;\; \cr
       \th \, \( \, [\, x - x_1 \, ]\, [\, x-x_2 \, ] \, \)
                & \quad \mbox{\frm in I} \;\;\, . \cr } \right.
\ee
The function $f_2$ now reads
\be   \label{B8}
  f_2 = {\o^2 -q^2 \0 q^2} \lk \o x - x^2
       - { 4 \mu^2 q^2 + (\o^2 - q^2 )^2 \0 4 (\o^2 - q^2 ) }
       \, \rk \;\; ,
\ee
and the expression for $\Im m \sum_i $ so far obtained is
\be   \label{B9}
  \Im m \Sigma_i = {T \0 8 \pi q} \int_\mu^\infty \!\! dx \,
        {1 \0 x} \, f_i \, \sgn (\o - x ) \, \Theta
        \; - \; ( \o \to - \o ) \;\; .
\ee
Clearly, the integrals (\ref{B9}) can be evaluated analytically, 
once the range of the parameters
is specified. Consider region III and assume $\o > 0$. Then
$\mu < x_1 < x_2 < \o$. Hence, $\sgn (\o - x) = +1$ and the term
$-\, (\o \to - \o)$ vanishes. Thus, in region III:
\be   \label{B10}
  \d \, \Im m \, \Sigma_i = {T\0 8\pi q} \lb
  \begin{array}{ll} \dis {\cal J} &  \quad i=1  \\
    \dis {\o^2 -q^2 \0 q^2} \( {q \O \0 2}
    - {4 \mu^2 q^2 + (\o^2 - q^2 )^2 \0 4 (\o^2 - q^2)}
    \, {\cal J} \, \)    &  \quad i=2
  \end{array}  \right.
\ee
with ${\cal J} = \ln \,\vert\, (\o^2 + q \O ) /
( \o^2 - q \O ) \,\vert\,  $ as given in (\ref{6J})
in the main text. Of course, in region III 
neither the prefix $\d$ nor the absolute value in the argument
of the logarithm are necessary. See however below.

In region I and for $\o > 0$ we have $x_1 < \mu < x_2$ and
$\sgn (\o -x) = -1$. Hence, in the first term of (\ref{B9}), the
$x$-integration runs from $x_2$ to $\infty$. The second term, i.e.
$- \, (\o \to - \o )$, needs one more position: $\overline{x}_1
= - {\o \0 2 } + {q \0 2 \o } \O\,$. Note that
$0 < \mu < \overline{x}_1 < x_2\,$. Also, $f_2$ might be split
into its even and odd part with respect to $\o$. In
region I and for positive $\o$ we obtain
\be   \label{B11}
 \Im m \Sigma_2 = {T\0 8\pi q}
   \( \int_{\overline{x}_1}^{x_2} \!\! dx \, {1 \0 x} \,
          f_2^{\hbox{{\frm even}}} - {\o^2 - q^2 \0 q^2}
      \, \o \, \lk \int_{\overline{x}_1}^\infty \!\! dx +
      \int_{x_2}^\infty \!\! dx \, \rk  \) \;\; .
\ee
Obviously, this expression needs subtraction of the hard
leading-order imaginary part, which is in fact non-zero in
region I. This amounts to subtracting $2 \int_0^\infty \!\! dx$
from the two diverging integrals in (\ref{B11}). The result agrees
precisely with (\ref{B10}), but the prefix $\d$ and the absolute
value in ${\cal J}$ are no longer superfluous. For $i=1$ there is
only the first integral in (\ref{B11}). For an immediate check,
whether the above partial results are correctly used in the main
text, the reader may take the imaginary parts of (\ref{2Psubg}),
(\ref{2Psubell}), insert (\ref{B10}) and compare with (\ref{6res_g})
and (\ref{6res_l}).

We turn to the real part. It has to be determined from the
dispersion relation (\ref{6disp}). But instead of directly
treating the corresponding integrals (a torture) there is a
pleasant way by guessing and using analytical properties.
First of all, we put a $\d$ in front of both sides of
(\ref{6disp}) to indicate subtraction of leading-order terms. The
integration over $t$ runs from $-\infty$ to $-\wu {4\mu^2 + q^2}$,
from $-q$ to $q$ and from $\wu {4\mu^2 + q^2}$ to $\infty$. This
fact may be dealt with by including the step function
\be  \label{B12}
   \Theta^\prime \equiv \th \( \, \lk \, q^2-t^2 \,\rk
       \lk \, 4 \mu^2 + q^2 - t^2 \,\rk  \,\) \;\; .
\ee
Let us start with the case $i=1$, i.e. with $f = 1$ in (\ref{6disp}).
The imaginary part is an odd function of $t$. To see this explicitly,
and by using (\ref{B10}), (\ref{6J}), (\ref{6Om}), we write
(\ref{6disp}) as
\be   \label{B13}
        \d \Re e \Sigma_1
      = {T \0 8\pi^2 q} \int \! dt \, { 1 \0 t-\o} \,
        L(t) \, \Theta^\prime  \;\; ,
\ee
with
\be   \label{B14}
  L(t) \equiv  \ln \, \left| \, {t^2 + qtW(t) \0 t^2 - qtW(t) }
  \; \right|  \quad \mbox{and} \quad
  W(t) \equiv \wu {{4\mu^2 + q^2 - t^2 \0 q^2 - t^2}} \;\; .
\ee
Principal values are understood whenever a real variable runs
over a pole. In order to get rid of the absolute value signs in
(\ref{B14}) we rewrite $L$ as
\be  \label{B15}
  L(t) = {1 \0 2} \ln \(
         {4\mu^2 + (tW+q)^2 \0 4 \mu^2 + (tW-q)^2 } \) \;\; .
\ee
Next we tackle with the step function in (\ref{B13}). Note that
$\Re e W(t)$ vanishes automatically in the unwanted region
$q^2 < t^2 < 4\mu^2+q^2$. The same holds true for $\Re e \, g(t)$
with
\be   \label{B16}
   g(z) = \ln \( {2\mu - izW(z) - iq \0 2\mu - izW(z) + iq }
          \) \;\; .
\ee
Moreover, in the other regions $\Re e \, g(t)$ agrees
with (\ref{B15}). Thus, in all regions
\be  \label{B17}
   L(t) \, \Theta^\prime \, = \Re e \, g(t) \, \equiv g_1  \;\; .
\ee
At this point we recall that $\Re e \Sigma_i$ and $\Im m \Sigma_i$
are originally defined by approaching the real axis from the upper
half complex plane (UHP). Note that, with $\e$ a positive
infinitesimal, $W(t+i\e )$ turns into $+i\vert W(t) \vert$
when entering region II from larger as well as from lower $t$. Both
the functions $W(z)$ and $g(z)$ have cuts on the real axis, but
they are analytic in the UHP. Since, in addition, $g(z)$
behaves as $1/z$ at $z \to \infty$, we may state
\be  \label{B18}
 {1 \0 2\pi i} \int_{{\cal C}} \! dz^\prime \,
 {1 \0 z^\prime - z} \, g(z^\prime ) = g(z) \;\; ,
\ee
where ${\cal C}$ surrounds the UHP counterclockwise.
Through $z \to t+i\e \,$ and $\, g(t+i\e )
\equiv g_1 + i g_2\,$, (\ref{B18}) tells us the common dispersion
relations
\be  \label{B19}
 g_2(t) = - {1 \0 \pi} \int \! dt^\prime
    { 1 \0 t^\prime - t} \, g_1(t^\prime ) \quad , \quad
 g_1(t) =  {1 \0 \pi} \int \! dt^\prime
    { 1 \0 t^\prime - t} \, g_2(t^\prime ) \;\; .
\ee
Combining the left of these equations with (\ref{B13}), (\ref{B17})
and (\ref{B16}), we have
\be   \label{B20}
  \Re e \Sigma_1 = - {T \0 8\pi q} \, g_2 (\o +i\e )
                 =  {T \0 8\pi q} \, {\cal R}
\ee
with ${\cal R}$ given by (\ref{6R}). The procedure is unique.
One can now verify $\d \P _g$ as given in (\ref{6res_g}).

In treating the case $i=2$ we follow the same lines of reasoning.
$\d \Re e \Sigma_2$ is given by the integral (\ref{B13})
but with $L(t)$ replaced by the function
\be  \label{B21}
 M(t) = {t^2 - q^2 \0 2 q} \, t \, W(t) -
 { 4 \mu^2 q^2 + (t^2 - q^2 )^2  \0 4 q^2 } L(t) \;\; .
\ee
To restore convergence of (\ref{B13}) we subtract (and add) on both
sides the same expression taken at $\mu = 0$, i.e. we split
\be  \label{B22}
   \d \Re e \Sigma_2 = \d_\mu \Re e \Sigma_2
  + \d \Re e \sum \D_0^- \D_0 f_2
\ee
with  $\d_\mu \Re e \Sigma_2 \equiv \Re e \Sigma_2 - \Re e
\Sigma_2^{\mu=0}$, and treat the second term of (\ref{B22}) at the
end. This decomposition was used recently in the QCD case (\S 4 of
\cite{HS}) with the terms called 'one-loop soft' and 'one-loop hard',
respectively. We observe that $\d_\mu M(t) \, \Theta^\prime$ is the
real part $\Re e \, g(\o + i \e )$ of the following complex function:
\bea  \label{B23}
  g(z) & = & (z^2-q^2) {z \0 2q} \lk W(z) - 1 \,\rk
         - { 4\mu^2 q^2 + (z^2-q^2)^2 \0 4q^2 }
        \,\ln \( {2\mu - izW(z) - iq \0 2\mu - izW(z) + iq }
        \)   \nonu \\
       & & {} + \, { (z^2-q^2)^2 \0 4q^2 } \,\ln \(
      {z+q+i\e \0 z-q+i\e } \)
      - i (z^2-q^2) \, {\mu \0 q} \;\; ,
\eea
where the last term, which obviously is not fixed by $g_1$, has been
determined from the requirement $g(z) \sim 1/z$ at $z \to \infty$.
The relations (\ref{B18}) and (\ref{B19}) hold true again, and thus
\be  \label{B24}
 \d_\mu \, \Re e \, \Sigma_2 = {T \0 8\pi^2 q}
  \int \! dt \, {1 \0 t-\o } \, g_1 (t)  = - {T \0 8 \pi q}
  \, g_2(\o + i \e ) \;\; .
\ee
Note that $\o W(\o + i \e )$ is equal to the quantity ${\cal E}$,
(\ref{6E}), used in the main text. With $g_2(\o + i \e )$ taken from
(\ref{B23}) we end up with
\be  \label{B25}   \hspace{-.3cm}
   \d_\mu \Re e \Sigma_2  = {T\0 8\pi q}
    \lk {\o^2-q^2 \0 2q}
    \(\, 2\mu - \Im m {\cal E} \,\)
    - {4\mu^2q^2+(\o^2-q^2)^2 \0 4q^2} {\cal R}
    + { (\o^2-q^2)^2 \0 4q^2 } \pi \Theta_I \rk \;\; ,
\ee
where $\Theta_I = \th (q^2-\o^2)$ restricts the last term to
region I.

It remains to study the second term of (\ref{B22}). Its imaginary
part may be read off from (\ref{B10}),
\be  \label{B26}
  \d \Im m \sum \D_0^- \D_0 f_2 = {T \0 8\pi q}
  {\o^2-q^2 \0 4q^2} \lk 2q\o- (\o^2 - q^2 )
  \ln \,\left| {\o + q \0 \o - q} \right| \,\,\,
      \rk \;\; ,
\ee
and the appropriate, in the UHP analytic function is
\be  \label{B27}
  g(z) = (z^2-q^2) {z\0 2q} - {(z^2-q^2)^2 \0 4q^2}
         \,\ln \( {z+q-i\e \0 z-q-i\e }
         \) \; - {1 \0 3}\, q \,z  \;\; .
\ee
Again the last term (though in conflict with (\ref{B26})) has been
added by hand to make $g(z)$ convergent at large $z$. In reality
convergence is restored by the Bose function (which one could
reintroduce in (\ref{B9}) by $1/x \to n(x)/T\, $). However, such
details (if real on the real axis) do not influence the result, as
we shall take the imaginary part. In fact,
$\Im m \, g(\o + i \e )
= g_2 = - (\o^2-q^2) \, \pi \, \Theta_I \,$.
Consequently, when the second term of (\ref{B22}) is included, the
last term in (\ref{B25}) simply drops out.

The results (\ref{6res_l}) to (\ref{6res_t}) in the main text, which
we have also derived by a much more laborious direct calculation,
are now readily verified.

%
%
\let\dq=\thq  \renewcommand{\theequation}{C.\dq}
\setcounter{equation}{0}

\parag {Appendix \ C}

In this Appendix, the ratio $\Re e \P_\ell / Q^2 $ is calculated
at the light cone $\o =q$, in order to verify the statement
(\ref{pllcres}) in the main text. Here the decomposition of the
Braaten-Pisarski scheme into soft and hard loop momenta, where only
the former need resummation, fails. Instead, the resummed version
(\ref{2Psubell}) of the longitudinal polarization $\P_\ell$ must be
used throughout even though the leading contribution will be seen to
arise from hard internal momenta. Also, the Bose function cannot be
expanded.

The Bose function may be reintroduced by the replacement
$T/x \to n(x)$ in the formulas of the preceding Appendix B.
The latest opportunity for doing so is at (\ref{B9}):
\bea  \label{C1}
 \Im m \,\Sigma_i & = & {1 \0 8\pi q} \int_\mu^\infty \! dx \, n(x)
        \lk h_2(\o ) - h_2 (-\o ) \rk  \\ \label{C2}
  & & \mbox{with} \quad  h_2(\o )
       = f_i (\o ) \,\sgn (\o -x) \,\Theta
\eea
and $f_1=1$. $f_2$ and $\Theta$ are given by (\ref{B8}) and
(\ref{B5}), respectively. The two sums $\Sigma_i$, which constitute
$\P_\ell\,$, are defined in (\ref{B1}). If we are able to grasp the
analytical function $h(z)$, which has $\Im m \, h(\o+i\e )
= h_2(\o )$ and is convergent at large $z$, we may proceed as in
Appendix B, but this time inside the real $x$-integration. Indeed,
we find
\be \label{C3}
 h(z) = f_i (z) {1\0\pi} \ln
     \( {z^2-q^2-2xz-2kq \0 z^2-q^2-2xz+2kq} \)
     - {k\0 \pi q }\, (z^2-q^2-2xz) \,\, \d_{\, i\, ,\, 2} \;\; ,
\ee
where $k=\wu {x^2-\mu^2}$. Only in the case $i=2$ there is a term to
be subtracted for convergence. Thus, with (\ref{B19}) and $h_1(\o )
\equiv \Re e\, h(\o + i \e ) $ we obtain the real parts
\be  \label{C4}
 \Re e\,\Sigma_i = {1 \0 8\pi q} \int_\mu^\infty \! dx \, n(x)
                  \lk h_1(\o ) + h_1 (-\o ) \rk \;\; .
\ee
One can now form the linear combination (\ref{2Psubell}) of the above
real parts, with all details filled in. The result is conveniently
written down with the integration variable $k$ instead of
$x=\wu {\mu^2+k^2} \equiv \wsc \;$:
\bea \label{C5}
 \Re e\,\P_\ell & = & (\o^2 -q^2) {e^2 \0 8 \pi^2 q^3}
   \int_0^\infty \! dk \, {k \0 \wsc } \, n(\wsc )
  \lk -(2\wsc - \o )^2 \ln \( {\o^2-q^2-2\wsc \o -2kq
          \0 \o^2-q^2-2\wsc \o + 2kq} \) \right. \nonu \\
  & & \left. -(2\wsc + \o )^2 \ln  \( {\o^2 - q^2 + 2\wsc \o
       - 2kq \0 \o^2-q^2+2\wsc \o + 2kq} \) -8kq \rk \;\; .
\eea

The expression (\ref{C5}) is the appropriate place for approaching
the light cone. After dividing both sides by $\o^2-q^2 = Q^2$,
(\ref{C5}) leads to
\be \label{C6}
 \lim_{\o \to q} \, {\Re e \P_\ell \0 Q^2} =
       {e^2 \0 \pi^2 q^2}\,  J \quad \mbox{with} \quad
    J \equiv \int_0^\infty \! dk \, {k \0 \wsc} \, n(\wsc )
    \lk -k + \wsc \,\ln \( {\wsc + k \0 \wsc -k} \) \rk \;\; .
\ee
To study the high temperature expansion of this integral $J$, we
subtract and add the large-$k$ limit of the integrand and write
\be  \label{C7}
 J = J_0 + \d J \quad \mbox{with} \quad
     J_0 = \int_0^\infty \! dk \, k \, n(k) \lk -1 + 2
           \ln \( {2k \0 \mu} \) \rk \;\; .
\ee
In the small difference term $\d J$ one can use $n(x) \approx T/x$,
so that it is readily evaluated to be $\d J = T\mu \pi / 2 $, which
leads to the contribution $ e^2 T \mu / 2\pi q^2 $ to $\Re e
\P_\ell / Q^2$. The leading term $J_0$ may be given the form
\be  \label{C8}
 J_0 = T^2 \lk - {\pi^2 \0 6} + {\pi^2 \0 3} \ln \({2T \0 \mu} \)
    + 2 \lim_{\e \to 0} {1 \0 \e } \int_0^\infty \!
    du {u^{1+\e }-u \0 e^u -1} \rk \;\; ,
\ee
where the last term brings in derivatives of the Riemann zeta and 
Gamma functions \cite{Abro}. Equation (\ref{pllcres}) is thus
obtained. Since accurate values of derivatives of the zeta function
are rarely found in tables, we note that
$ {\z^\prime (2) \0 \z (2)} = -0.569\,960\,993\ldots$.

We now turn to the imaginary part of the resummed polarization
$\P_\ell$ at the level of (\ref{C5}). It derives from (\ref{C1}),
(\ref{C2}) directly and is strictly zero in region II.
For $\o^2 < 4\mu^2+q^2$, i.e. in the
two regions I and II which have the light cone as the common border,
we find
\be  \label{C10} \hspace{-.1cm}
  \Im m\, \P_\ell(\o+i\e,q) = - Q^2 \, \th \( q^2-\o^2 \) \,
       {e^2 \0 2\pi q^3} \int_{x_0}^\infty \! dx \, x^2
       \lk
       n \( x- {\o\0 2} \) - n \( x+ {\o\0 2} \)  \rk \;\; . \;\;
\ee
Here, the effect of resummation is hidden in the lower endpoint
$x_0$ of the integration interval:
\be  \label{C11}
  x_0 = {1 \0 2}\, q\, \wu { 4\mu^2+q^2-\o^2 \0 q^2-\o^2 } \;\; .
\ee
Clearly, due to $\mu \neq 0$, $x_0$ comes down from infinity as $q$
becomes larger than $\o$, giving a smooth onset of the imaginary part
in region I. For $\o<q$ and $q-\o\ll e^2 q$,
\be \label{Cim}
  \Im m \P_\ell(\o+i\e,q) = {e^4 T^2 \0 8\pi}
       \exp \( - e \wu { q \0 8 (q-\o ) } \)  \;\; ,
\ee
so that the imaginary part sets in infinitely smoothly in fact.
Alarmingly, (\ref{Cim}) has the wrong sign to give rise to weak
damping. As it stands, it would give {\it anti-damping} as soon as
the light-cone is traversed into region I. The simple resolution of
this startling puzzle is that, in region I, $ \Im m \P_\ell(\o
-i\e,q) = -\Im m \P_\ell(\o+i\e,q)$. In region II, where the
next-to-leading order dispersion curve is still undamped,
higher-order corrections will generate positive damping. The
dispersion curve, when it is followed towards and through the
light-cone, crosses into region I with $Q_0 = \o(q) -i\g(q)$, so
that only the Riemann sheet where $\Im m \P_\ell < 0$ can be
accessed, which corresponds to Landau {\it damping} instead of
anti-damping. Consequently, for $(q-\o)/(eT) \ll e^2$ one has weakly
damped plasmon excitations with phase velocities just below 1,
whereas further down in region I these excitations quickly become
overdamped.

%
%
\let\dq=\thq  \renewcommand{\theequation}{D.\dq}
\setcounter{equation}{0}

\parag {Appendix \ D}

In this Appendix we add some details to the calculations presented
in Chap. 7.

First, as a consistency check we rederive Eqs.~(\ref{pir},\ref{piru})
from an equivalent Schwinger-Dyson equation, where the diagrams of
Fig.~7 are flipped so that the bare vertices are on the right side.
Since $\P\mn^{(R)}$ is an asymmetric combination of the real-time
components, this leads to a different starting point for
$\P\mn^{(1)R}$. In place of Eq.~(\ref{pir1}) we obtain
\bea \label{pirfl}  \hspace{-.5cm}
  \P\mn^{(1)R}(Q)&\equiv&\P\mn^{(1)11}(Q)+\P\mn^{(1)12}(Q) \nonu\\
  & = & {e\02}\int{d^4K\0(2\pi)^4}\Bigl\{
  \lk \G^{R(i)}_\mu +\G^{P(o)}_\mu \rk \D_R(Q+K)\D_R(K) \nonu\\
  & & {} + \lk \G^{R(o)}_\mu +\G^{P(i)}_\mu \rk \D_A(Q+K)
           \D_A(K) \nonu\\
  & & {} + \G^{R(\g)}_\mu \lk \D_R(Q+K) \D_P(K)
         + \D_P(Q+K) \D_A(K) \rk \Bigr\}(2K+Q)_\nu \;\; , \;\;
\eea
where we have additionally introduced
\be
  \G^{P(i)} = \sum_{a,b} \G^{baa} \;\; , \qquad
  \G^{P(o)} = \sum_{a,b} \G^{aab} \;\; .
\ee

The Ward identities that allowed us to solve $\G^{R(i)}$,
$\G^{R(o)}$, and $\G^{(P)}$ in terms of the corresponding
self-energies can be derived by subjecting the type-1 and type-2
field to the same gauge transformations \cite{LS2}. This would not
determine the vertices appearing above. Assuming, however, that the
theory is separately invariant under type-1 and type-2 gauge
transformations relates all components of the vertices to the
corresponding self-energy components. In analogy to (\ref{wardis})
we are then led to
\bea         \hspace{-.55cm}
  \G^{R(\g)}_\mu(K,K+Q) & = & -ie(2K+Q)_\mu \(
      1-{1\0 2KQ+Q^2}\lk \X_A(K)-\X_R(K+Q) \rk \) \; , \;  \\
              \hspace{-.55cm}
  \G^{P(i)}_\mu(K,K+Q) & = & -ie(2K+Q)_\mu
                         {1\0 2KQ + Q^2}\X_P(K+Q) \;\; , \\
              \hspace{-.55cm}
  \G^{P(o)}_\mu(K,K+Q) & = & ie(2K+Q)_\mu
  {1\0 2KQ + Q^2}\X_P(K) \;\; .
\eea
Inserting these into (\ref{pirfl}) and keeping only terms involving
the Bose distribution function exactly reproduces (\ref{piru}).

In the following we shall concentrate on $\P\mn^R$ in the form
(\ref{pir}), since there the contributions from the poles of the
Bose distribution functions that are proportional to $T^2$ or to $T$
cancel upon integration of the 3-momenta. Closing the contour around
the upper half plane, only the poles of $G_A(K)$ contribute, which
are located at $K_0=i\z\pm\w$ with $\w=\wu{k^2+\mu^2-\z^2}$.

We consider first $\P_{00}(Q_0,0)$, which when nonvanishing signals
the loss of gauge invariance. Always omitting temperature-independent
contributions, Eq.~(\ref{pir}) leads to
\bea
  \P_{00}(Q_0,0)& = & {} -{e^2\0Q_0+2i\z}\int{dk\,k^2\0\pi^2}\,
      \biggl\{ n(\wp)-n(\wm) \nonu\\
  & & {} - {i\z\0Q_0\w}(\wp+Q_0)\lk n(\wp+Q_0)-n(\wp) \rk \nonu\\
  & & {} + {i\z\0Q_0\w}(\wm-Q_0)\lk n(\wm-Q_0)-n(\wm) \rk \biggr\}
  \;\; .
\eea
Taylor-expanding the distribution functions and keeping only terms up
to $\z^3$, we have
\bea
  \P_{00}(Q_0,0) & = & {} -{e^2i\z\0Q_0+2i\z}\int{dk\,k^2\0\pi^2}
     \,\biggl\{ 2n'(\w)-\z^2n'''(\w) \nonu\\
  & & {} - {\wp+Q_0\0\w}\sum_{m=1}^\infty \lk n^{(m)} + i\z n^{(m+1)}
         - {\z^2\02} n^{(m+2)} \rk {Q_0^{m-1}\0m!} \nonu\\
  & & {} - {\wm-Q_0\0 \w}\sum_{m=1}^\infty
  \lk n^{(m)} - i\z n^{(m+1)} - {\z^2\02} n^{(m+2)} \rk
  {(-Q_0)^{m-1}\0 m!} \biggr\} \;\; .
\eea
The integrals associated with higher powers of $\z$ are increasingly
infrared-dominated, so that the terms neglected are down by powers
of $\z/\mu\sim e\ln(1/e)$. Collecting the various powers in $Q_0$,
we find for $Q_0\ll T$
\bea     \hspace{-.3cm}
  \P_{00}(Q_0,0) & = & {4i\z^3 e^2T\0 Q_0+2i\z}\int{dk\,k^2\0\pi^2}
     \sum_{m=1}^\infty \,\biggl\{  \nonu\\
  & & {} - {n Q_0^{2m-2}\0 (k^2+\mu^2)^{m+1}} + {(n+1)Q_0^{2m}\0
           (k^2+\mu^2)^{m+2}} + O(\z/\mu)\biggr\} \;\; . \;\;
\eea
This sum is telescoping --- all terms but the first cancel each
other, and (\ref{pooz3}) is readily obtained.

Considering next $\P\mn(0,q)$, we have
\bea
  \P_{00}(0,q) & = & {e^2\0\pi^2}\Re e \int dk\,k^2 {n(\wp)\0 \w}
       \biggl[ 1 \nonu\\
  & & {} + {k^2+\mu^2-2\z^2+2i\z\w\0kq}\ln{2kq-4i\z\w+q^2+4\z^2\0
  -2kq-4i\z\w+q^2+4\z^2} \biggr]\\
  \P_{ii}(0,q) & = & {} -{e^2\0\pi^2}\Re e \int dk\,k^2 {n(\wp)\0 \w}
   \biggl[ 1 \nonu\\
  & & {} - {4k^2+8i\z\w-q^2-8\z^2\04kq}\ln{2kq-4i\z\w+q^2+4\z^2\0
         - 2kq-4i\z\w+q^2+4\z^2} \biggr] \;\; .
\eea
In the limit $q\to0$ this simplifies to
\bea
  \P_{00}(0,q\to0) & = & {e^2\0\pi^2}\Re e \int dk \, k^2
        {in(\wp)\0\z} \nonu\\
  & = & {e^2\0\pi^2} \int dk\,k^2 \(-n'(\wu{k^2+\mu^2}) +
  {\z^2\06}n'''(\wu{k^2+\mu^2})+\ldots \) \;\; \quad \\
  \P_{ii}(0,q\to0) & = & {} -{e^2\0\pi^2} \int dk\,k^2 \(
  {\z^2\03}{k^4\0k^2+\mu^2}n'''(\wu{k^2+\mu^2})+\ldots\) \;\; ,
\eea
which yields Eqs.~(\ref{pz3}).

%
%
\newpage               \renewcommand{\section}{\paragraph}


\begin{thebibliography}{41}  \bigskip
\bibitem{Kapu} J. I. Kapusta, ``Finite-temerature field theory'',
   Cambridge University Press, Cambridge, 1989.
\bibitem{Linde} A. D. Linde, {\it Phys. Lett. B} {\bf 96} (1980),
   289; D. J. Gross, R. D. Pisarski and L. G. Yaffe, {\it Rev. Mod.
   Phys.} {\bf 53} (1981), 43.
\bibitem{samm}
   O. K. Kalashnikov and V. V. Klimov, {\it Sov. J. Nucl. Phys.}
   {\bf 31} (1980), 699;
   K. Kajantie and J. Kapusta, {\it Ann. Phys. (NY)} {\bf 160}
   (1985), 477;
   U. Heinz, K. Kajantje and T. Toimela, {\it Ann. Phys. (NY)}
   {\bf 176} (1987), 218;
   T. H. Hansson and I. Zahed, {\it Nucl. Phys. B} {\bf 292}
   (1987), 725;
   R. Kobes and G. Kunstatter, {\it Phys. Rev. Lett.} {\bf 61}
   (1988), 392;
   S. Nadkarni, {\it Phys. Rev. Lett.} {\bf 61} (1988), 396;
   U. Kraemmer, M. Kreuzer and A. Rebhan, {\it Ann. Phys. (NY)}
   {\bf 201} (1990), 223;
   M. Kreuzer, A. Rebhan and H. Schulz, {\it Phys. Lett. B}
   {\bf 244} (1990), 58.
\bibitem{BP} E. Braaten and R. D. Pisarski, {\it Nucl. Phys. B}
   {\bf 337} (1990), 569.
\bibitem{DJ} S. Weinberg, {\it Phys. Rev. D} {\bf 9} (1974), 2257;
   L. Dolan and R. Jackiw, {\it Phys. Rev. D} {\bf 9} (1974), 3320.
\bibitem{AE} P. Arnold and O. Espinosa, {\it Phys. Rev. D} {\bf 47}
   (1993), 3546.
\bibitem{BHW} W. Buchm\"uller, W. Helbig and D. Walliser, {\it Nucl.
   Phys. B} {\bf 407} (1993), 387;
   A. Hebecker, {\it Z. Phys. C} {\bf 60}, (1993), 271.
\bibitem{Damping}
   R. D. Pisarski, {\it Phys. Rev. Lett.} {\bf 63} (1989), 1129;
   E. Braaten and R. D. Pisarski, {\it Phys. Rev. D} {\bf 42}
   (1990), 2156; {\it ibid.} {\bf 46} (1992), 1829;
   R. Kobes, G. Kunstatter and K. Mak, {\it Phys. Rev. D} {\bf 45}
   (1992), 4632.
\bibitem{BMAR} C. P. Burgess and A. L. Marini, {\it Phys. Rev. D}
   {\bf 45} (1992), R17;
   A. Rebhan, {\it Phys. Rev. D} {\bf 46} (1992), 482.
\bibitem{HS} H. Schulz, {\it Nucl. Phys. B} {\bf 413} (1994), 353.
\bibitem{AKR}  A. K. Rebhan, {\it Phys. Rev. D} {\bf 48} (1993),
   3967.
\bibitem{MC} M. E. Carrington, {\it Phys. Rev. D} {\bf 48} (1993),
   3836.
\bibitem{IZ} C. Itzykson and J. Zuber, ``Quantum Field Theory'',
   McGraw-Hill, New York, 1985.
\bibitem{LW} P. Landsman and Ch. van Weert, {\it Phys. Reports}
   {\bf 145} (1987), 141.
\bibitem{Silin} V. P. Silin, {\it ZhETF} {\bf 38} (1960), 1577.
\bibitem{Fradkin} E. S. Fradkin, {\it Proc. of the Lebedev Institute}
   {\bf 29} (1965), 6.
\bibitem{KalKl} O. K. Kalashnikov and V. V. Klimov, {\it Yad. Fiz.}
   {\bf 33} (1981), 848
   [{\it Sov. J. Nucl. Phys.} {\bf 33} (1981), 443].
\bibitem{Weldon} H. A. Weldon, {\it Phys. Rev. D} {\bf 26} (1982),
   1394.
\bibitem{GMB} M. Gell-Mann and K. A. Brueckner, {\it Phys. Rev.}
   {\bf 106} (1957), 364.
\bibitem{FT} J. Frenkel and J. C. Taylor, {\it Nucl. Phys. B}
   {\bf 334} (1990), 199.
\bibitem{BPward} E. Braaten and R. D. Pisarski, {\it Nucl. Phys. B}
   {\bf 339} (1990), 310.
\bibitem{eff} J. C. Taylor and S. M. Wong, {\it Nucl. Phys. B}
   {\bf 346} (1990), 115.
\bibitem{shortcut} R. D. Pisarski, in: ``From fundamental fields
   to nuclear phenomena'', eds. J. A. McNeil and C. E. Price,
   World Scientific Publ. Co., 1991.
\bibitem{Parwani} R. R. Parwani, {\it Phys. Rev. D} {\bf 45} (1992),
   4695.
\bibitem{EHKT} H.-Th. Elze, U. Heinz, K. Kajantie and T. Toimela,
   Z. Phys. C {\bf 37} (1988), 305.
\bibitem{Weldonmis} H. A. Weldon, {\it Phys. Rev. D} {\bf 47} (1993),
   594.
\bibitem{Pisres} R. D. Pisarski, {\it Physica A} {\bf 158} (1989),
   146.
\bibitem{SU} V. P. Silin and V. N. Ursov, {\it Sov. Phys. - Lebedev
   Inst. Rep.} {\bf 5} (1988), 43.
\bibitem{LS} V. V. Lebedev and A. V. Smilga, {\it Ann. Phys.}
   {\bf 202} (1990), 229.
\bibitem{Tsyt} V. N. Tsytovich, {\it Sov. Phys. JETP} {\bf 13}
   (1961), 1249.
\bibitem{BPS} R. Baier, S. Peign\'e and D. Schiff, preprint
   LPTHE-Orsay 93/46, BI-TP 93/55 (1993) (hep-ph/9311329), 
   to appear in {\it Z. Phys. C.}
\bibitem{LS2} V. V. Lebedev and A. V. Smilga, {\it Physica A}
   {\bf181} (1992), 187.
\bibitem{APG} T. Altherr, E. Petitgirard and T. del Rio
   Gaztelurrutia, {\it Phys. Rev. D} {\bf 47} (1993), 703.
\bibitem{BNN} R. Baier, H. Nakkagawa and A. Ni\'egawa, {\it Can. J.
   Phys.} {\bf 71} (1993), 205.
\bibitem{fermdamp} A. V. Smilga, preprint BUTP-92/39 (1992);
   R. Pisarski, {\it Phys. Rev. D} {\bf 47} (1993), 5589;
   S. Peign\'e, E. Pilon and D. Schiff, {\it Z. Phys. C}
   {\bf 60} (1993), 455.
\bibitem{SchwK} L. V. Keldysh, {\it Sov. Phys. JETP} {\bf 20}
   (1964), 1018;
   E. M. Lifshitz and L. P. Pitaevsky, ``Physical Kinetics'',
   Pergamon Press, Oxford, 1981.
\bibitem{RA} R. Kobes, {\it Phys. Rev. D} {\bf 42} (1990), 562;
   P. Aurenche and T. Becherrawy, {\it Nucl. Phys. B} {\bf 379}
   (1992), 259;
   M. A. van Eijck and Ch. G. van Weert, {\it Phys. Lett. B}
   {\bf 278} (1992), 305.
\bibitem{KaKli} O. K. Kalashnikov and V. V. Klimov, {\it Phys.
   Lett. B} {\bf 95} (1980), 423.
\bibitem{AS} T. Altherr and D. Seibert, {\it Phys. Lett. B}
   {\bf 313} (1993), 149; {\bf 316} (1993), 633 (E).
\bibitem{Niev}J. F. Nieves and P. B. Pal, ``The zero-momentum limit
   of thermal Green functions'', preprint DOE-ER 40757-040 (1994)
   (hep-ph/9402290).
\bibitem{rel} J. Frenkel and J. C. Taylor, {\it Nucl. Phys. B}
   {\bf 374} (1992), 156.
\bibitem{BPeff} E. Braaten and R. D. Pisarski, {\it Phys. Rev. D}
   {\bf 45} (1992), R1827.
\bibitem{Abro} M. Abramovitz and I.A. Stegun, ``Handbook of
   Mathematical Functions'', Dover Publications, New York, 1970.
\end{thebibliography}
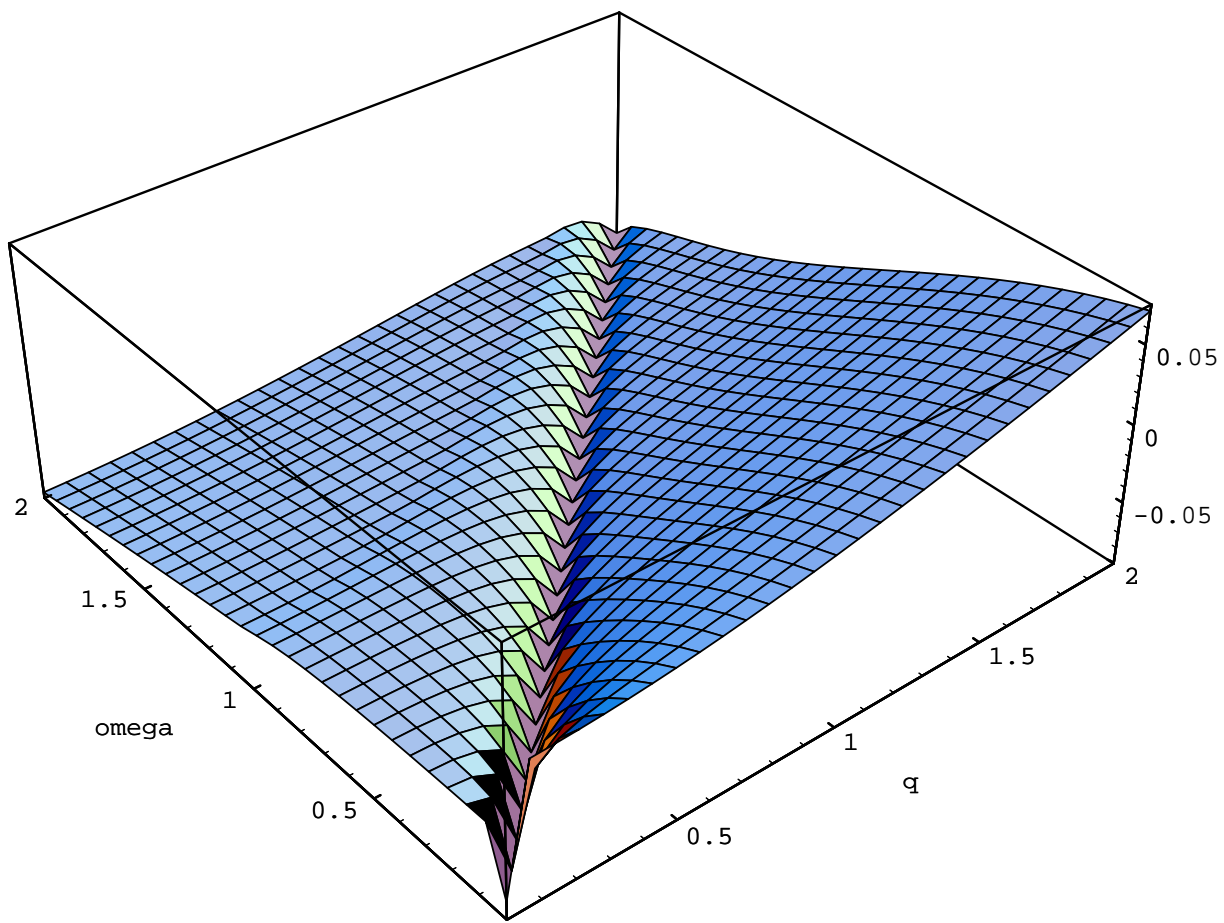
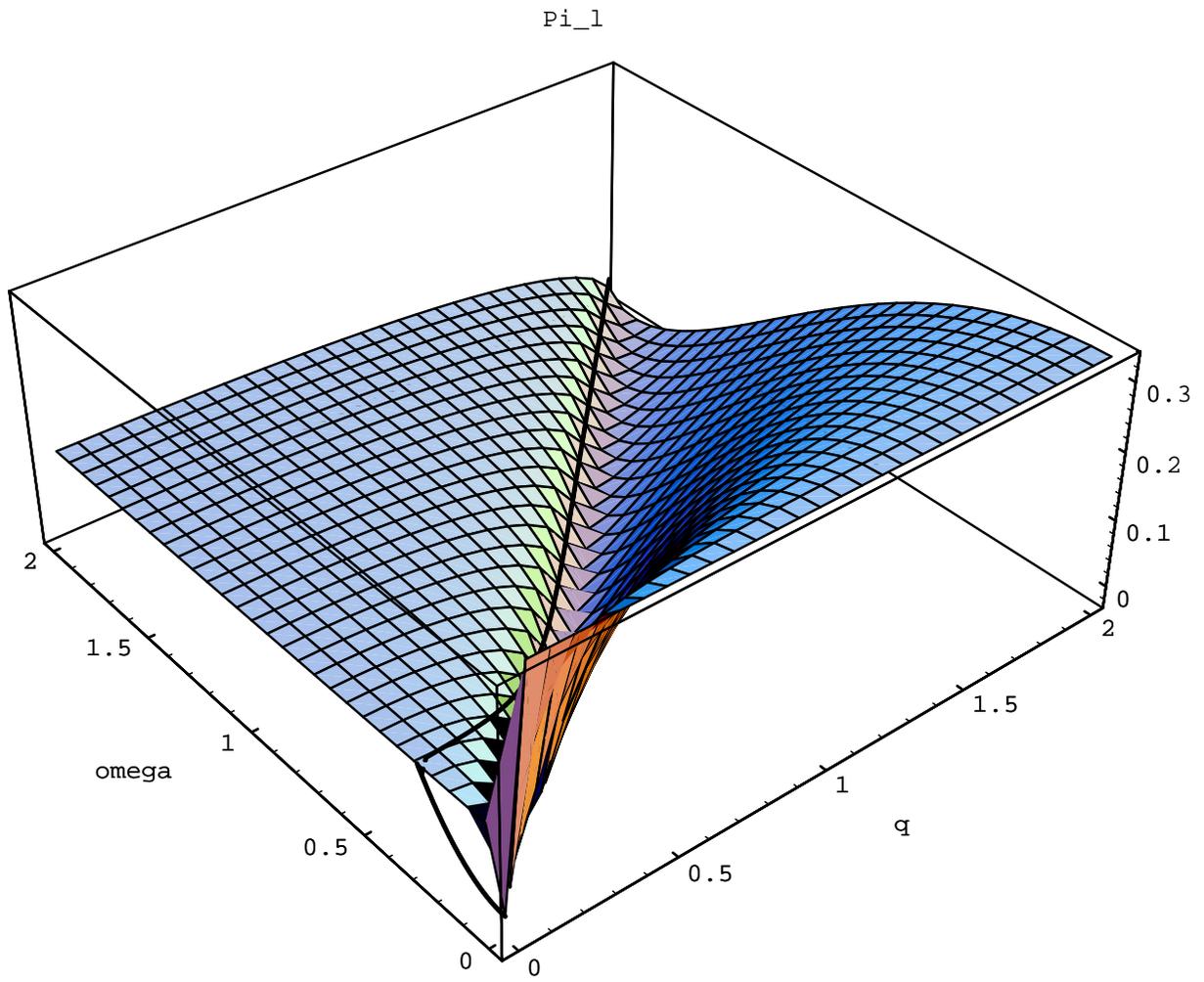
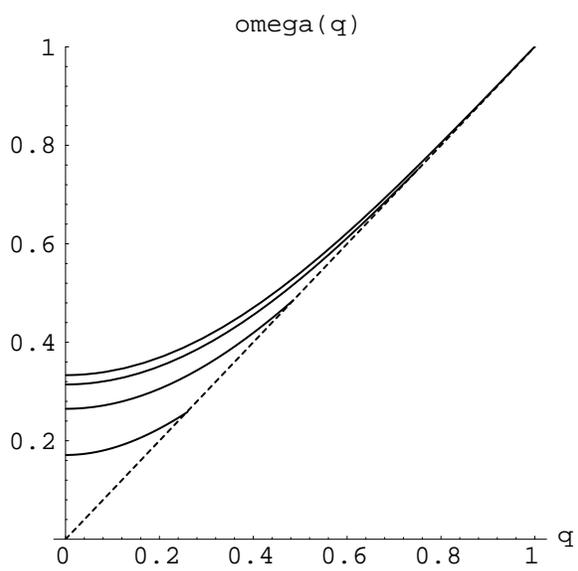
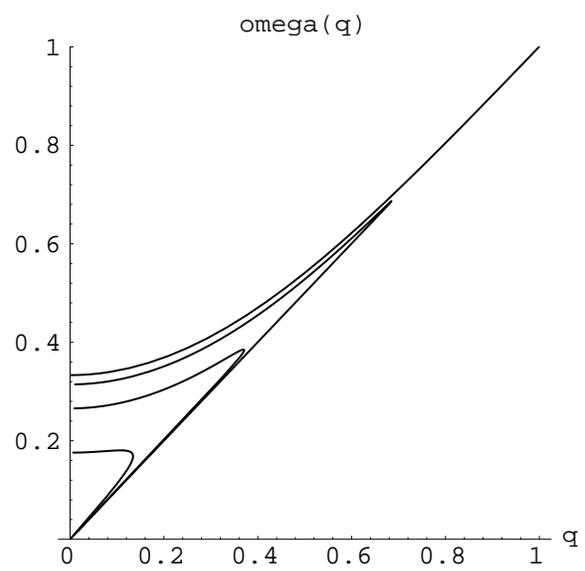
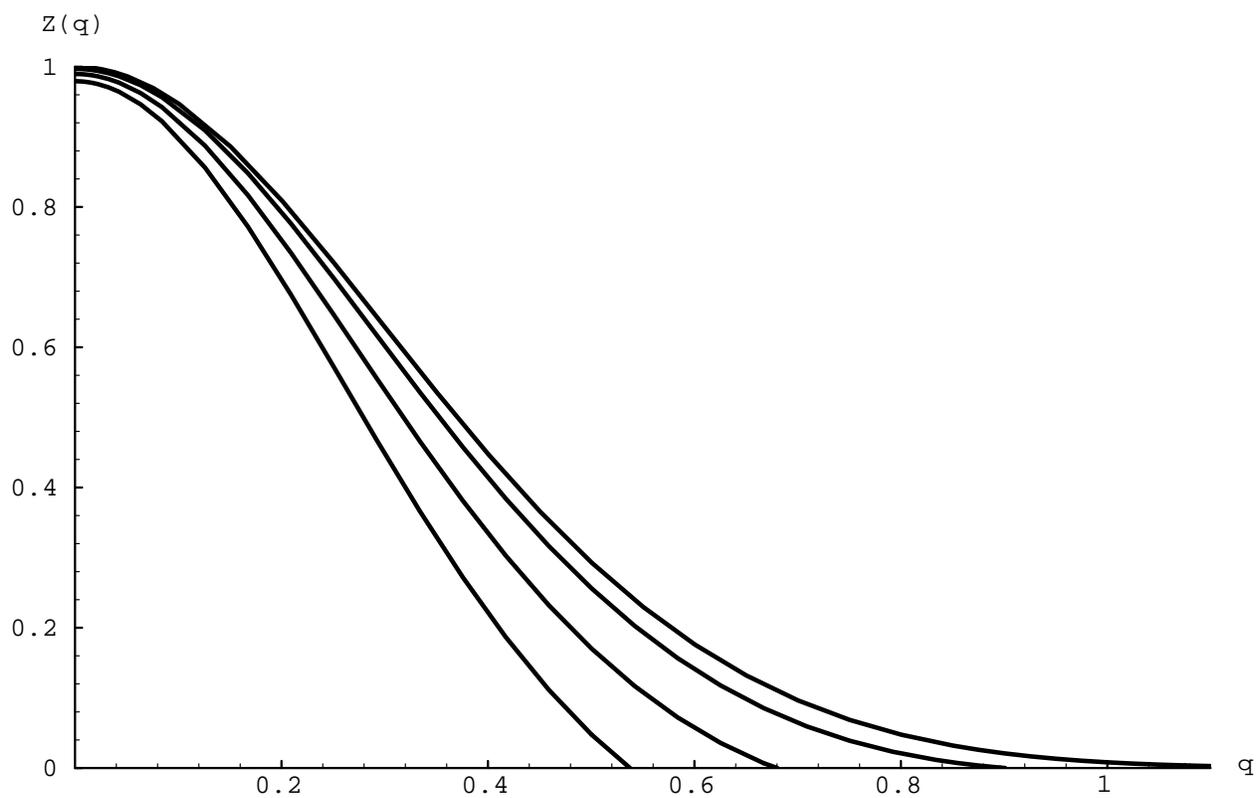
\end{document}